\documentclass[twoside]{ae100prg}

\usepackage{graphicx}
\usepackage[breaklinks]{hyperref}
\usepackage{booktabs}
\usepackage{bm}

\begin{document}
\title{The General Relativistic Two Body Problem and the Effective One Body Formalism}


\author{Thibault Damour$^1$}

\address{$^1$ Institut des Hautes \'Etudes Scientifiques, 35 route de Chartres, 91440 Bures-sur-Yvette, France}

\email{damour@ihes.fr}

\begin{abstract}
A new analytical approach to the motion and radiation of (comparable mass) binary systems has been introduced in 1999 under the name of Effective One Body (EOB) formalism. We review the basic elements of this formalism, and discuss some of its recent developments. Several recent comparisons between EOB predictions and Numerical Relativity (NR) simulations have shown the aptitude of the EOB formalism to provide accurate descriptions of the dynamics and radiation of various binary systems (comprising black holes or neutron stars) in regimes that are inaccessible to other analytical approaches (such as the last orbits and the merger of comparable mass black holes). In synergy with NR simulations, post-Newtonian (PN) theory and Gravitational Self-Force (GSF) computations, the EOB formalism is likely to provide an efficient way of computing the very many accurate template waveforms that are needed for Gravitational Wave (GW) data analysis purposes.
\end{abstract}

\section{Introduction}

The general relativistic $N$-body problem has been investigated from the early days of Einstein's gravitation theory (and even earlier, because it was already tackled by Johannes Droste within the framework of the 1913 Einstein-Grossmann ``Entwurf'' theory). Here, we shall focus on the general relativistic two-body problem. This problem has been the subject of many investigations within the post-Newtonian (PN) formalism, since the pioneering works of Einstein (1915; when $m_1 \ll m_2$), Lorentz and Droste (1917), Levi-Civita (1937) and Einstein, Infeld and Hoffmann (1938). [see, \textit{e.g.}, \cite{Damour1987} for a review and references to the early literature.] For many years, the first post-Newtonian (1PN) approximation (\textit{i.e.} the inclusion of the leading-order relativistic corrections, proportional to $(v/c)^2$ or $GM / c^2 r$, to the Newtonian equations of motion) appeared as being accurate enough for applying Einstein's theory to known binary systems. The situation changed in the mid 1970's with the discovery of the Hulse-Taylor binary pulsar PSR~$1913+16$. The need to compare the accurate observations of this system (by Taylor and collaborators) to the predictions of Einstein's theory motivated the development of improved relativistic theories of binary systems, applicable to strongly self-gravitating bodies, and including terms up to the $2.5$PN approximation (\textit{i.e.} $O [(v/c)^5]$ beyond Newton). [See \cite{DamourLH} and references therein.] The situation has again changed recently with the development of interferometric gravitational wave (GW) detectors, and the prospect of detecting the GW's emitted during the last orbits and the coalescence of binary systems made of black holes or neutron stars. The latter prospect motivated the development (or improvement) of several different methods of computing the motion and radiation of binary systems.

\medskip

First, this motivated pushing PN calculations of the dynamics of binary systems to the $3$PN level \cite{Jaranowski1998,Blanchet2001,Damour2001,Blanchet2004,Itoh2003}, with inclusion of $3.5$PN radiation-reaction terms \cite{Pati2002,Konigsdorffer2003,Nissanke2005}. 
Second, this motivated the development of new, accurate GW generation formalisms, notably the Blanchet-Damour-Iyer (matched) ``multipolar post-Minkowskian'' formalism \cite{Blanchet1986,Blanchet1989,Damour1991,DamourIyer1991,Blanchet1992,Blanchet1995} and the ``direct integration of the relaxed Einstein's equations'' formalism of Will and collaborators \cite{Will1996,Will1999,Will2000}, which extended previous work by Epstein and Wagoner \cite{Epstein1975} and Thorne \cite{Thorne1980}. These GW generation formalisms allowed one to compute emitted gravitational waveforms with an unprecedented PN accuracy\footnote{For gravitational waveforms, one conventionally defines the PN accuracy as the \textit{fractional} PN accuracy with respect to the leading-order, $O(c^{-5})$, quadrupolar emission. \textit{E.g.}, a $1$PN-accurate waveform retains next-to-leading order terms, \textit{i.e.} terms smaller than the quadrupolar waveform by a factor $O(c^{-2})$.}. After the $1$PN correction to the waveform \cite{Epstein1975,Wagoner1976,Blanchet1989}, there is a $1.5$PN ``tail'' (\textit{i.e.} hereditary) correction \cite{Blanchet1992,Wiseman1993,Blanchet1993}, then a ``direct'' $2$PN term \cite{BDIWW1995,BDI1995,Will1996}, followed by higher-order corrections \cite{Blanchet1998,Blanchet2002,Blanchet2005,BDEI2004,BDEI2005,LBlanchet1998,Berti2007,Kidder2008,BFIS2008}. [See \cite{LBlanchet2002} for a detailed account and more references.] Parallely to these improved PN computations of the GW emission of comparable-mass systems (with $m_1 \sim m_2$), other authors developed the analytical theory of the GW emission of extreme mass-ratio systems (with $m_1 \ll m_2$): see Refs.~\cite{Poisson1993,Sasaki1994,Tagoshi1994,Tanaka1996} and the review of Sasaki and Tagoshi \cite{Sasaki2003}.

Some of the PN calculations of the dynamics, and/or GW emission, of comparable-mass systems
have been recently (re)done (e.g. the 3PN dynamics~\cite{Foffa2011}) by using a somewhat different formalism, dubbed ``effective field theory''~\cite{Goldberger:2004jt}. Let us, however, note that most of the technical aspects of the
effective-field-theory approach had already been introduced and used before. For instance: (i) Ref.~\cite{Damour:1992we}
discussed the (Fokker) two-body effective action due to the exchange of a linear field (of spin $s=0,1$ and $2$) ;
(ii) Ref.~\cite{Damour:1995kt} explicitly
discussed the representation (and  computation) of the (nonlinear) effective two-body action in terms of  Feynman-like
diagrams (made of concatenated propagators and vertices);
(iii)  The appendix A of Ref.~\cite{Damour:1998jk} discussed
finite-size effects in terms of nonminimal worldline couplings in the effective action; (iv) The (quantum field theory)
 technique of dimensional regularization (together with a diagrammatic analysis of  ultraviolet divergences)
 was crucially used to derive the 3PN dynamics in Refs.~\cite{Damour2001,Blanchet2004},   and 3PN radiation
 in Ref.~\cite{BDEI2005};
and (v) The exponential parametrization of the metric (which suppresses the leading-order gravitational cubic vertex) had been introduced
in Ref.~\cite{Blanchet1989} and then standardly used in many PN works.
 It is, however, possible that the more
systematic (and automatized) implementation of such diagrammatic methods, together with the tapping of  standard techniques for computing Feynman graphs
 (as exemplified in ~\cite{Foffa2011})
may allow one to be more efficient
in computing higher-order processes, or, at least, to open new ways of understanding them 
(see, in this respect, Ref.~\cite{Goldberger:2009qd}).

\medskip

Separately from these purely analytical approaches to the motion and radiation of binary systems, which have been developed since the early days of Einstein's theory, Numerical Relativity (NR) simulations of Einstein's equations have relatively recently (2005) succeeded (after more than thirty years of developmental progress) to stably evolve binary systems made of comparable mass black holes \cite{Pretorius2005,Campanelli2006,Baker2006,Boyle2007}. This has led to an explosion of works exploring many different aspects of strong-field dynamics in General Relativity, such as spin effects, recoil, relaxation of the deformed horizon formed during the coalescence of two black holes to a stationary Kerr black hole, high-velocity encounters, etc.; see \cite{Pretorius2007} for a review. In addition, recently developed codes now allow one to accurately study the orbital dynamics, and the coalescence of binary neutron stars. Much physics remains to be explored in these systems, especially during and after the merger of the neutron stars (which involves a much more complex physics than the pure-gravity merger of two black holes).

\medskip

Recently, a new source of information on the general relativistic two-body problem has opened: gravitational self-force (GSF) theory. This approach goes one step beyond the test-particle approximation (already used by Einstein in 1915) by taking into account self-field effects that modify the leading-order geodetic motion of a small mass $m_1$ moving in the background geometry generated by a large mass $m_2$. After some ground work (notably by DeWitt and Brehme) in the 1960's, GSF theory has recently undergone rapid developments (mixing theoretical and numerical methods) and can now yield numerical results that yield access to new information on strong-field dynamics in the extreme mass-ratio limit $m_1 \ll m_2$. See Ref.~\cite{Barack2009} for a review.

\medskip

Each of the approaches to the two-body problem mentioned so far, PN theory, NR simulations and GSF theory, have their advantages and their drawbacks. It has become recently clear that the best way to meet the challenge of accurately computing the gravitational waveforms (depending on several continuous parameters) that are needed for a successful detection and data analysis of GW signals in the upcoming LIGO/Virgo/GEO/$\ldots$ network of GW detectors is to combine knowledge from all the available approximation methods: PN, NR and GSF. Several ways of doing so are a priori possible. For instance, one could try to directly combine PN-computed waveforms (approximately valid for large enough separations, say $r \gtrsim 10 \, G (m_1 + m_2)/c^2$) with NR waveforms (computed with initial separations $r_0 > 10 \, G(m_1 + m_2)/c^2$ and evolved up to merger and ringdown). However, this method still requires too much computational time, and is likely to lead to waveforms of rather poor accuracy, see, \textit{e.g.}, \cite{Hannam2008,MHannam2008,MacDonald:2012mp}.

\medskip

On the other hand, five years before NR succeeded in simulating the late inspiral and the coalescence of binary black holes, a new approach to the two-body problem was proposed: the Effective One Body (EOB) formalism \cite{Buonanno1999,Buonanno2000,DJS2000,TDamour2001}. The basic aim of the EOB formalism is to provide an analytical description of both the motion and the radiation of coalescing binary systems over the entire merger process, from the early inspiral, right through the plunge, merger and final ringdown. As early as 2000~\cite{Buonanno2000} this method made several quantitative 
and qualitative predictions concerning the dynamics of the coalescence, and 
the corresponding GW radiation, notably: (i) a blurred transition from
inspiral to a `plunge' that is just a smooth continuation of the inspiral, 
(ii) a sharp transition, around the merger of the black holes, between a 
continued inspiral and a ring-down signal, and (iii) estimates of the radiated 
energy and of the spin of the final black hole. In addition, the effects of 
the individual spins of the black holes were investigated within the 
EOB~\cite{TDamour2001,Buonanno2006} and were shown to lead to a larger 
energy release for spins parallel to the orbital angular momentum, and to a 
dimensionless rotation parameter $J/E^2$ always smaller than unity at the 
end of the inspiral (so that a Kerr black hole can form right after the 
inspiral phase). All those predictions have been broadly confirmed by the 
results of the recent numerical simulations performed by several independent 
groups (for a review of numerical relativity results and references see~\cite{Pretorius2007}). 
Note that, in spite of the high computer power used in NR simulations, the calculation, checking and processing of one sufficiently long waveform (corresponding to specific values of the many continuous parameters describing the two arbitrary masses, the initial spin vectors, and other initial data) takes on the order of one month. This 
is a very strong argument for developing analytical models of waveforms.

\section{EOB description of the conservative dynamics of two body systems}

Before reviewing some of the technical aspects of the EOB method, let us
indicate the historical roots of this method. First, we note that 
the EOB approach comprises three, rather separate, ingredients:
\begin{enumerate}
\item{ a description of the conservative (Hamiltonian) part of the dynamics of two bodies;}
\item{ an expression for the radiation-reaction part of the dynamics;}
\item{ a description of the GW waveform emitted by a coalescing binary system.}
\end{enumerate}

\medskip

For each one of these ingredients, the essential inputs that are used in EOB 
works are high-order post-Newtonian (PN) expanded results which have 
been obtained by many years of work, by many researchers (see references
above). However, one of the key ideas in the EOB philosophy is to avoid using
PN results in their original ``Taylor-expanded'' form (\textit{i.e.} $c_0 + c_1 \,
v/c + c_2 \, v^2/c^2 + c_3 \, v^3/c^3 + \cdots + c_n \, v^n/c^n)$, but to use them instead in some \textit{resummed} form (\textit{i.e.} some non-polynomial function of 
$v/c$, defined so as to incorporate some of the expected non-perturbative 
features of the exact result). The basic ideas and techniques for resumming 
each ingredient of the EOB are different and have different historical roots. 

\medskip

Concerning the first ingredient, \textit{i.e.} the EOB Hamiltonian, it was inspired 
by an approach to electromagnetically interacting quantum two-body systems 
introduced by Br\'ezin, Itzykson and Zinn-Justin~\cite{Brezin1970}.

\medskip

The resummation of the second ingredient, \textit{i.e.} the EOB radiation-reaction 
force ${\mathcal F}$, was initially inspired by the Pad\'e resummation of 
the flux function introduced by Damour, Iyer and Sathyaprakash~\cite{Damour1998}.
More recently, a new and more sophisticated resummation technique for the 
radiation reaction force ${\mathcal F}$ has been introduced by 
Damour and Nagar~\cite{Damour2009,DamourNagar2009}. 

\medskip

As for the third ingredient, \textit{i.e.} the EOB description of the waveform 
emitted by a coalescing black hole binary, it was mainly inspired by the 
work of Davis, Ruffini and Tiomno~\cite{Davis1972} which discovered 
the transition between the plunge signal and a ringing tail when a particle 
falls into a black hole. [Additional motivation for the EOB treatment of 
the transition from plunge to ring-down came from work on the, 
so-called, ``close limit approximation''~\cite{Price1994}.] In addition, a very efficient
resummation of the waveform has been introduced by Damour, Iyer and Nagar~\cite{Damour2007,Damour2008,Damour2009}.  It will be discussed in detail below.

\medskip

Within the usual PN formalism, the conservative dynamics of a two-body system is currently fully known up to the $3$PN level \cite{Jaranowski1998,Blanchet2001,Damour2001,Blanchet2004,Itoh2003,Foffa2011} (see below for the partial knowledge beyond the $3$PN level). Going to the center of mass of the system $({\bm p}_1 + {\bm p}_2 = 0)$, the $3$PN-accurate Hamiltonian (in Arnowitt-Deser-Misner-type coordinates) describing the relative motion, ${\bm q} = {\bm q}_1 - {\bm q}_2$, ${\bm p} = {\bm p}_1 = - {\bm p}_2$, has the structure
\begin{equation}
\label{eq7.1}
H_{\rm 3PN}^{\rm relative} ({\bm q} , {\bm p}) = H_0 ({\bm q} , 
{\bm p}) + \frac{1}{c^2} \, H_2 ({\bm q} , {\bm p}) + \frac{1}{c^4} \, H_4 ({\bm q} , {\bm p}) + \frac1{c^6} \, H_6 ({\bm q} , {\bm p}) \, ,
\end{equation}
where 
\begin{equation}
\label{HNewton}
H_0 ({\bm q} , {\bm p}) = \frac{1}{2\mu} \, {\bm p}^2 - \frac{GM\mu}
{\vert {\bm q} \vert} \, ,
\end{equation}
with 
\begin{equation}
M \equiv m_1 + m_2 \quad \mbox{and} \quad \mu \equiv m_1 \, m_2 / M \, ,
\end{equation}
 corresponds to the Newtonian approximation to the relative motion, while $H_2$ 
describes 1PN corrections, $H_4$ 2PN ones and $H_6$ 3PN ones. In terms of the rescaled variables ${\bm q}' \equiv {\bm q} / GM$, ${\bm p}' \equiv {\bm p} / \mu$, the explicit form (after dropping the primes for readability) of the 3PN-accurate rescaled Hamiltonian $\widehat H \equiv H / \mu$ reads \cite{Damour2000,TDamour2000,Damour2001}
\begin{equation}
\label{4.28a}
\widehat H_N ({\bm q} , {\bm p}) = \frac{{\bm p}^2}2 - \frac1q \, ,
\end{equation}

\begin{equation}
\label{4.28b}
\widehat H_{\rm 1PN} ({\bm q} , {\bm p}) = \frac18 (3\nu - 1)({\bm p}^2)^2 - \frac12 [(3+\nu) {\bm p}^2 + \nu ({\bm n} \cdot {\bm p})^2] \frac1q + \frac1{2q^2} \, ,
\end{equation}

$$
\widehat H_{\rm 2PN} ({\bm q} , {\bm p}) = \frac1{16} (1-5\nu + 5\nu^2) ({\bm p}^2)^3 
$$
$$
+ \frac18 [(5-20\nu - 3\nu^2)({\bm p}^2)^2 -2\nu^2 ({\bm n} \cdot {\bm p})^2 {\bm p}^2 - 3\nu^2 ({\bm n} \cdot {\bm p})^4] \frac1q
$$
\begin{equation}
\label{4.28c}
+ \frac12 [(5 + 8\nu) {\bm p}^2 + 3\nu ({\bm n} \cdot {\bm p})^2] \frac1{q^2} - \frac14 (1+3\nu) \frac1{q^3} \, ,
\end{equation}

$$
\widehat H_{\rm 3PN} ({\bm q} , {\bm p}) = \frac1{128} (-5 + 35\nu - 70\nu^2 + 35\nu^3)({\bm p}^2)^4
$$
$$
+ \frac1{16} [(-7 + 42\nu - 53\nu^2 - 5\nu^3)({\bm p}^2)^3 + (2-3\nu)\nu^2 ({\bm n} \cdot {\bm p})^2({\bm p}^2)^2 
$$
$$
+ \, 3 (1-\nu) \nu^2 ({\bm n} \cdot {\bm p})^4 {\bm p}^2 - 5\nu^3 ({\bm n} \cdot {\bm p})^6] \frac1q
$$
$$
+ \left[ \frac1{16} (-27 + 136\nu + 109\nu^2)({\bm p}^2)^2 + \frac1{16} (17+30\nu) \nu ({\bm n} \cdot {\bm p})^2 {\bm p}^2 \right.
$$
$$
\left. + \frac1{12} (5+43\nu)\nu ({\bm n} \cdot {\bm p})^4 \right] \frac1{q^2}
$$
$$
+ \left\{ \left[ -\frac{25}8 + \left( \frac1{64} \pi^2 - \frac{335}{48} \right) \nu - \frac{23}8 \nu^2 \right] {\bm p}^2  \right.
$$
$$
\left. + \left( -\frac{85}{16} - \frac3{64} \pi^2 - \frac74 \nu \right) \nu ({\bm n} \cdot {\bm p})^2 \right\} \frac1{q^3}
$$
\begin{equation}
\label{4.28d}
+ \left[ \frac18 + \left( \frac{109}{12} - \frac{21}{32} \pi^2 \right) \nu \right] \frac1{q^4} \, .
\end{equation}

In these formulas $\nu$ denotes the symmetric mass ratio:
\begin{equation}
\nu \equiv \frac{\mu}{M} \equiv \frac{m_1 \, m_2}{(m_1 + m_2)^2} \, .
\end{equation}
The dimensionless parameter $\nu$ varies between $0$ (extreme mass ratio case) and $\frac14$ (equal mass case) and plays the r\^ole of a deformation parameter away from the test-mass limit.

\medskip

It is well known that, at the  Newtonian approximation, $H_0 ({\bm q} , {\bm p})$ can be thought of as  describing a `test particle' of mass $\mu$ orbiting around an `external 
mass' $GM$. The EOB approach is a \textit{general relativistic generalization} 
of this fact. It consists in looking for an `effective external spacetime geometry'
 $g_{\mu\nu}^{\rm eff} (x^{\lambda} ; GM,\nu)$ such that the geodesic dynamics of 
a `test particle' of mass $\mu$ within $g_{\mu\nu}^{\rm eff} (x^{\lambda} , 
GM,\nu)$ is \textit{equivalent}  (when expanded in powers of $1/c^2$) to the
original, relative PN-expanded dynamics (\ref{eq7.1}). 

\medskip

Let us explain the idea, proposed in \cite{Buonanno1999}, for establishing 
a `dictionary' between the real relative-motion dynamics, (\ref{eq7.1}), and 
the dynamics of an `effective' particle of mass $\mu$ moving in 
$g_{\mu\nu}^{\rm eff} (x^{\lambda} , GM,\nu)$. The idea consists in `thinking 
quantum mechanically'\footnote{This is related to an idea emphasized many 
times by John Archibald Wheeler: quantum mechanics can often help us in 
going to the essence of classical mechanics.}. Instead of thinking in terms 
of a classical Hamiltonian, $H({\bm q}, {\bm p})$ 
(such as $H_{\rm 3PN}^{\rm relative}$, Eq.~(\ref{eq7.1})), and of its classical 
bound orbits, we can think in terms of the quantized energy levels $E(n,\ell)$ 
of the quantum bound states of the Hamiltonian operator $H (\hat{\bm q} ,\hat{\bm p})$. These energy levels will depend on two (integer valued) quantum numbers 
$n$ and $\ell$. Here (for a spherically symmetric interaction, as appropriate 
to $H^{\rm relative}$), $\ell$ parametrizes the total orbital angular 
momentum (${\bm L}^2 = \ell (\ell + 1) \, \hbar^2$), while $n$ represents 
the `principal quantum number' $n = \ell + n_r + 1$, where $n_r$ 
(the `radial quantum number') denotes the number of nodes in the radial 
wave function. The third `magnetic quantum number' $m$ 
(with $-\ell \leq m \leq \ell$) does not enter the energy levels because 
of the spherical symmetry of the two-body interaction (in the center of 
mass frame). For instance, the non-relativistic Newton interaction Eq.~(\ref{HNewton}) 
gives rise to the well-known result
\begin{equation}
\label{eqn2}
E_0 (n,\ell) = - \frac{1}{2} \, \mu \left(\frac{GM\mu}{n \, \hbar} \right)^2 \, ,
\end{equation}
which depends only on $n$ (this is the famous Coulomb degeneracy). 
When considering the PN corrections to $H_0$, as in Eq.~(\ref{eq7.1}), 
one gets a more complicated expression of the form
$$
E_{\rm 3PN}^{\rm relative} (n,\ell) = - \frac{1}{2} \mu  
\frac{\alpha^2}{n^2} \biggl[ 1 + 
\frac{\alpha^2}{c^2} \left( \frac{c_{11}}{n\ell} + \frac{c_{20}}{n^2} \right)
$$
\begin{equation}
\label{eqn3}
+ \frac{\alpha^4}{c^4} \left( \frac{c_{13}}{n\ell^3} + \frac{c_{22}}{n^2 \ell^2} + \frac{c_{31}}{n^3 \ell} + \frac{c_{40}}{n^4} \right) + \frac{\alpha^6}{c^6} \left( \frac{c_{15}}{n\ell^5} + \ldots + \frac{c_{60}}{n^6} \right)\biggl] \, ,
\end{equation}
where we have set $\alpha \equiv GM\mu / \hbar = G \, m_1 \, m_2 / \hbar$, 
and where we consider, for simplicity, the (quasi-classical) limit 
where $n$ and $\ell$ are large numbers. The 2PN-accurate version of Eq.~(\ref{eqn3}) 
had been derived by Damour and Sch\"afer \cite{Damour1988} 
as early as 1988 while its 3PN-accurate version was derived by Damour, Jaranowski and Sch\"afer in 1999 \cite{Damour2000}. The dimensionless coefficients $c_{pq}$ are 
functions of the symmetric mass ratio $\nu \equiv \mu / M$, for 
instance $c_{40} = \frac{1}{8} (145 - 15\nu + \nu^2)$. 
In classical mechanics (\textit{i.e.} for large $n$ and $\ell$), it 
is called the `Delaunay Hamiltonian', \textit{i.e.} the Hamiltonian 
expressed in terms of the 
\textit{action variables}\footnote{We consider, for simplicity, 
`equatorial' motions with $m=\ell$, \textit{i.e.}, classically, 
$\theta = \frac{\pi}{2}$.} $J = \ell \hbar 
= \frac{1}{2\pi} \oint p_{\varphi} \,d\varphi$, 
and $N = n \hbar = I_r + J$, with $I_r = \frac{1}{2\pi} \oint p_r \, dr$.

\medskip

The energy levels (\ref{eqn3}) encode, in a \textit{gauge-invariant} way, 
the 3PN-accurate relative dynamics of a `real' binary. Let us 
now consider an auxiliary problem: the `effective' dynamics of 
one body, of mass $\mu$, following (modulo the $Q$ term discussed below) a geodesic in some $\nu$-dependent `effective external' 
(spherically symmetric) metric\footnote{It is convenient to write 
the `effective metric' in Schwarzschild-like coordinates. 
Note that the effective radial coordinate $R$ differs from 
the two-body ADM-coordinate relative distance 
$R^{\rm ADM} = \vert {\bm q} \vert$. The transformation 
between the two coordinate systems has been determined 
in Refs.~\cite{Buonanno1999,DJS2000}.}
\begin{equation}
\label{eq7.2}
g_{\mu\nu}^{\rm eff} \, dx^{\mu} \, dx^{\nu} = - A(R;\nu) \, c^2 \, d T^2 + B(R;\nu) \, d R^2 + R^2 (d\theta^2 + \sin^2 \theta \, d \varphi^2) \, .
\end{equation}
Here, the \textit{a priori unknown} metric functions $A(R;\nu)$ and $B(R;\nu)$ 
will be constructed in the form of expansions in $GM/c^2 R$:
\begin{eqnarray}
\label{eqn4}
A(R;\nu) &= &1 + \widetilde a_1 \, \frac{GM}{c^2 R} + \widetilde a_2 \left( \frac{GM}{c^2 R} \right)^2 + \widetilde a_3 \left( \frac{GM}{c^2 R} \right)^3 +  \widetilde a_4 \left( \frac{GM}{c^2 R} \right)^4 + \cdots \, ; \nonumber \\
B(R;\nu) &= &1 + \widetilde b_1 \, \frac{GM}{c^2 R} + \widetilde b_2 \left( \frac{GM}{c^2 R} \right)^2 + b_3 \left( \frac{GM}{c^2 R} \right)^3  + \cdots \, ,
\end{eqnarray}
where the dimensionless coefficients $\widetilde a_n , \widetilde b_n$ depend on $\nu$. From the Newtonian limit, it is clear that we should set $\widetilde a_1 = -2$. In addition, as $\nu$ can be viewed as a deformation parameter away from the test-mass limit, we require that the effective metric (\ref{eq7.2}) tend to the Schwarzschild metric (of mass $M$) as $\nu \to 0$, \textit{i.e.} that
$$
A(R;\nu=0)=1-2GM/c^2R = B^{-1} (R;\nu = 0) \, .
$$

\medskip

Let us now require that the dynamics of the ``one body'' $\mu$ within the effective metric $g_{\mu\nu}^{\rm eff}$ be described by an ``effective'' mass-shell condition of the form
$$
g_{\rm eff}^{\mu\nu} \, p_{\mu}^{\rm eff} \, p_{\nu}^{\rm eff} + \mu^2 \, c^2 + Q(p_{\mu}^{\rm eff}) = 0 \, ,
$$
where $Q(p)$ is (at least) \textit{quartic} in $p$. Then 
by solving (by separation of variables) the corresponding `effective'
 Hamilton-Jacobi equation
$$
g_{\rm eff}^{\mu\nu} \, \frac{\partial S_{\rm eff}}{\partial x^{\mu}} \, \frac{\partial S_{\rm eff}}{\partial x^{\nu}} + \mu^2 c^2 + Q \left( \frac{\partial S}{\partial x^{\mu}} \right) = 0 \, ,
$$
\begin{equation}
\label{eqn5}
S_{\rm eff} = - {\mathcal E}_{\rm eff} \, t + J_{\rm eff} \, \varphi + S_{\rm eff} (R) \, ,
\end{equation}
one can straightforwardly compute (in the quasi-classical, large quantum
numbers limit) the effective Delaunay 
Hamiltonian ${\mathcal E}_{\rm eff} (N_{\rm eff} , J_{\rm eff})$, 
with $N_{\rm eff} = n_{\rm eff} \, \hbar$, $J_{\rm eff} = \ell_{\rm eff} \,
\hbar$ (where $N_{\rm eff} = J_{\rm eff} + I_R^{\rm eff}$, with $I_R^{\rm eff}
= \frac{1}{2\pi} \oint p_R^{\rm eff} 
\, dR$, $P_R^{\rm eff} = \partial S_{\rm eff} (R) / dR$). 
This  yields a result of the form
\begin{eqnarray}
{\mathcal E}_{\rm eff} (n_{\rm eff},\ell_{\rm eff}) &= &\mu c^2 - \frac{1}{2} \,
\mu \ \frac{\alpha^2}{n_{\rm eff}^2}
\biggl[ 1 + \frac{\alpha^2}{c^2} \left( \frac{c_{11}^{\rm eff}}{n_{\rm eff}
    \ell_{\rm eff}} + \frac{c_{20}^{\rm eff}}{n_{\rm eff}^2} \right)\nonumber\\
&+& \frac{\alpha^4}{c^4} \left( \frac{c_{13}^{\rm eff}}{n_{\rm eff} \ell_{\rm eff}^3} 
+ \frac{c_{22}^{\rm eff}}{n_{\rm eff}^2 \ell_{\rm eff}^2} + \frac{c_{31}^{\rm eff}}{n_{\rm eff}^3 \ell_{\rm eff}} 
+ \frac{c_{40}^{\rm eff}}{n_{\rm eff}^4} \right) \nonumber \\
& + &\frac{\alpha^6}{c^6} \left( \frac{c_{15}^{\rm eff}}{n_{\rm eff} \ell_{\rm eff}^5} + \ldots + \frac{c_{60}^{\rm eff}}{n_{\rm eff}^6} \right) \biggl] ,
\label{eqn6}
\end{eqnarray}
where the dimensionless coefficients $c_{pq}^{\rm eff}$ are now 
functions of the unknown coefficients $\widetilde a_n , \widetilde b_n$ entering the 
looked for `external' metric coefficients (\ref{eqn4}).

\medskip

\begin{figure}[t]
\begin{center}
\includegraphics[height=5cm]{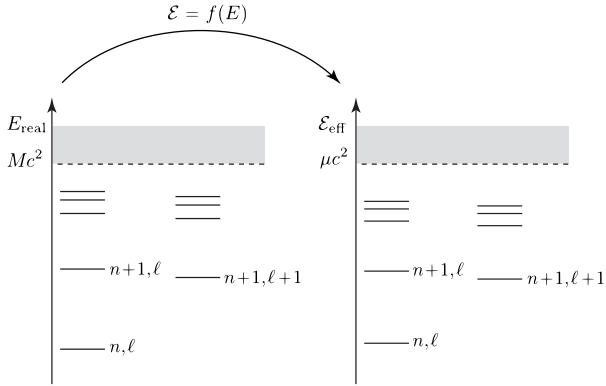}
\caption{\label{fig:1}Sketch of the correspondence between the quantized energy levels of the real and effective conservative dynamics. $n$ denotes the `principal 
quantum number' ($n = n_r + \ell + 1$, with $n_r = 0,1,\ldots$ denoting the
number of nodes in the radial function), while $\ell$ denotes the (relative) orbital 
angular momentum $({\bm L}^2 = \ell (\ell + 1) \, \hbar^2)$. Though the EOB 
method is purely classical, it is conceptually useful to think in terms of 
the underlying (Bohr-Sommerfeld) quantization conditions of the action 
variables $I_R$ and $J$ to motivate the identification between $n$ and 
$\ell$ in the two dynamics.}
\end{center}
\end{figure}

At this stage, one needs to define a `dictionary' between the real (relative) two-body dynamics, summarized in Eq.~(\ref{eqn3}), and the effective one-body one, summarized in Eq.~(\ref{eqn6}). As, on both sides, quantum mechanics tells us that the action variables are quantized in integers ($N_{\rm real} = n \hbar$, $N_{\rm eff} = n_{\rm eff} \hbar$, etc.) it is most natural to identify $n=n_{\rm eff}$ and $\ell = \ell_{\rm eff}$. One then still 
needs a rule for relating the two different energies $E_{\rm real}^{\rm relative}$ 
and ${\mathcal E}_{\rm eff}$. Ref.~\cite{Buonanno1999} proposed to look for a 
general map between the real energy levels and the effective ones (which, as seen 
when comparing (\ref{eqn3}) and (\ref{eqn6}), cannot be directly identified because they do not include the same rest-mass contribution\footnote{Indeed $E_{\rm real}^{\rm total} = Mc^2 + E_{\rm real}^{\rm relative} = Mc^2 + \mbox{Newtonian terms} + {\rm 1PN} / c^2 + \cdots$, while ${\mathcal E}_{\rm effective} = \mu c^2 + N + {\rm 1PN} / c^2 +\cdots$.}), namely 
$$
\frac{{\mathcal E}_{\rm eff}}{\mu c^2} - 1 = f \left( \frac{E_{\rm real}^{\rm relative}}{\mu c^2} \right) = \frac{E_{\rm real}^{\rm relative}}{\mu c^2} \biggl( 1 + \alpha_1 \, \frac{E_{\rm real}^{\rm relative}}{\mu c^2} + \alpha_2 \left( \frac{E_{\rm real}^{\rm relative}}{\mu c^2} \right)^2 
$$
\begin{equation}
\label{eqn7}
+ \, \alpha_3 \left( \frac{E_{\rm real}^{\rm relative}}{\mu c^2} \right)^3 + \ldots \biggl)  \, .
\end{equation}
The `correspondence' between the real and effective energy levels is illustrated in Fig.~\ref{fig:1}.

\medskip

Finally, identifying ${\mathcal E}_{\rm eff} (n,\ell) / \mu c^2$ to $1+f (E_{\rm real}^{\rm relative} (n,\ell) / \mu c^2)$ yields a system of equations for determining the unknown EOB coefficients $\widetilde a_n , \widetilde b_n , \alpha_n$, as well as the three coefficients $z_1 , z_2 , z_3$ parametrizing a general $3$PN-level quartic mass-shell deformation:
$$
Q_{\rm 3PN} (p) = \frac1{c^6} \, \frac1{\mu^2} \left(\frac{GM}{R} \right)^2 \left[ z_1 \, {\bm p}^4 + z_2 \, {\bm p}^2 ({\bm n} \cdot {\bm p})^2 + z_3 ({\bm n} \cdot {\bm p})^4 \right] \, .
$$
[The need for introducing a quartic mass-shell deformation $Q$ only arises at the 3PN level.]

\medskip

The above system of equations for $\widetilde a_n , \widetilde b_n , \alpha_n$ (and $z_i$ at 3PN) was studied at the 2PN level in Ref.~\cite{Buonanno1999}, and at the 3PN level in Ref.~\cite{DJS2000}. At the 2PN level it was found that, if one further imposes the natural condition $\widetilde b_1 = +2$ (so that the linearized effective metric coincides with the linearized Schwarzschild metric with 
mass $M = m_1 + m_2$), there exists a \textit{unique} solution for the remaining five unknown coefficients $\widetilde a_2 , \widetilde a_3 , \widetilde b_2 , \alpha_1$ and $\alpha_2$. 
This solution is very simple:
\begin{equation}
\label{eqn8}
\widetilde a_2 = 0  \, , \quad \widetilde a_3 = 2 \nu \, , \quad \widetilde b_2 = 4 - 6 \nu \, , \quad \alpha_1 = \frac{\nu}{2} \, , \quad \alpha_2 = 0 \, .
\end{equation}
At the 3PN level, it was found that the system of equations is consistent, and underdetermined in that the general solution can be parametrized by the arbitrary values of $z_1$ and $z_2$. It was then argued that it is natural to impose the simplifying requirements $z_1 = 0 = z_2$, so that $Q$ is proportional to the fourth power of the (effective) radial momentum $p_r$. With these conditions, the solution is unique at the 3PN level, and is still remarkably simple, namely
$$
\widetilde a_4 = a_4 \, \nu \, , \ \widetilde d_3 = 2(3\nu - 26) \nu \, , \ \alpha_3 = 0 \, , \ z_3 = 2(4-3\nu)\nu \, .
$$
Here, $a_4$ denotes the number
\begin{equation}
\label{eq7.6}
a_4 = \frac{94}{3} - \frac{41}{32} \, \pi^2 \simeq 18.6879027 \ ,
\end{equation}
while $\widetilde d_3$ denotes the coefficient of $(GM/c^2 R)^3$ in the PN expansion of the combined metric coefficient
$$
D(R) \equiv A(R) \, B(R) \, .
$$
Replacing $B(R)$ by $D(R)$ is convenient because (as was mentioned above), in the test-mass limit $\nu \to 0$, the effective metric must reduce to the Schwarzschild metric, namely
$$
A(R;\nu = 0) = B^{-1} (R;\nu=0) = 1 - 2 \left( \frac{GM}{c^2 R} \right) \, ,
$$
so that 
$$
D(R;\nu = 0) = 1 \, .
$$

\medskip

The final result is that the three EOB potentials $A,D,Q$ describing the 3PN two-body dynamics are given by the following very simple results. In terms of the EOB ``gravitational potential''
$$
u \equiv \frac{GM}{c^2 R} \, ,
$$
\begin{equation}
\label{eq7.5}
A_{\rm 3PN} (R) = 1-2u + 2 \, \nu \, u^3 + a_4 \, \nu \, u^4 \, ,
\end{equation}
\begin{equation}
\label{eq:D}
D_{\rm 3PN}(R)\equiv(A(R)B(R))_{\rm 3PN} = 1-6\nu u^2+2(3\nu-26)\nu u^3 \ ,
\end{equation}
\begin{equation}
\label{Q3PN}
Q_{\rm 3PN} ({\bm q} , {\bm p}) = \frac1{c^2} \, 2 (4-3\nu)\nu \ u^2 \ \frac{p_r^4}{\mu^2} \, .
\end{equation}

\medskip

In addition, the map between the (real) center-of-mass energy of the binary system $E_{\rm real}^{\rm relative} = H^{\rm relative} = {\mathcal E}_{\rm relative}^{\rm tot} - Mc^2$ and the effective one ${\mathcal E}_{\rm eff}$ is found to have the very simple (but non trivial) form
\begin{equation}
\label{eqn9}
\frac{{\mathcal E}_{\rm eff}}{\mu c^2} = 1 + \frac{E_{\rm real}^{\rm relative}}{\mu c^2} 
\left( 1 + \frac{\nu}{2} \, \frac{E_{\rm real}^{\rm relative}}{\mu c^2}\right) 
= \frac{s - m_1^2 \, c^4 - m_2^2 \, c^4}{2 \, m_1 \, m_2 \, c^4}
\end{equation}
where $s = ({\mathcal E}_{\rm real}^{\rm tot})^2 \equiv (M c^2 + E_{\rm  real}^{\rm relative})^2$  is Mandelstam's invariant $s=-(p_1+p_2)^2$. 

\medskip

It is truly remarkable that the EOB formalism succeeds in \textit{condensing} the complicated, original 3PN Hamiltonian, Eqs.~(\ref{4.28a})--(\ref{4.28d}), into the very simple potentials $A, D$ and $Q$ displayed above, together with the simple energy map Eq.~(\ref{eqn9}). For instance, at the 1PN level, the already somewhat involved Lorentz-Droste-Einstein-Infeld-Hoffmann 1PN dynamics (Eqs.~(\ref{4.28a}) and (\ref{4.28b})) is simply described, within the EOB formalism, as a test particle of mass $\mu$ moving in an external Schwarzschild background of mass $M = m_1 + m_2$, together with the (crucial but quite simple) energy transformation (\ref{eqn9}). [Indeed, the $\nu$-dependent corrections to $A$ and $D$ start only at the 2PN level.] At the 2PN level, the seven rather complicated $\nu$-dependent coefficients of $\widehat H_{\rm 2PN} ({\bm q} , {\bm p})$, Eq.~(\ref{4.28c}), get condensed into the two very simple additional contributions $+ \, 2 \nu u^3$ in $A(u)$, and $- \, 6 \nu u^2$ in $D(u)$. At the 3PN level, the eleven quite complicated $\nu$-dependent coefficients of $\widehat H_{\rm 3PN}$, Eq.~(\ref{4.28d}), get condensed into only three simple contributions: $+ \, a_4 \nu u^4$ in $A(u)$, $+ \, 2 (3\nu - 26) \nu u^3$ in $D(u)$, and $Q_{\rm 3PN}$ given by Eq.~(\ref{Q3PN}). This simplicity of the EOB results is not only due to the reformulation of the PN-expanded Hamiltonian into an effective dynamics. Indeed, the $A$-potential 
happens to be much simpler that it could a priori have been: (i) as already noted it is not modified at the 1PN level, while one would a priori expect to have found a 1PN potential $A_{\rm 1PN} (u) = 1-2u + \nu a_2 u^2$ with some non zero $a_2$; and (ii) there are striking cancellations taking place in the calculation of the 2PN and 3PN coefficients $\widetilde a_2 (\nu)$ and $\widetilde a_3 (\nu)$, which were a priori of the form $\widetilde a_2 (\nu) = a_2 \nu + a'_2 \nu^2$, and $\widetilde a_3 (\nu) = a_3 \nu + a'_3 \nu^2 + a''_3 \nu^3$, but for which the $\nu$-nonlinear contributions $a'_2 \nu^2 , a'_3 \nu^2$ and $a''_3 \nu^3$ precisely cancelled out.

\medskip

The fact that the 3PN coefficient $a_4$ in the crucial 
`effective radial potential' $A_{\rm 3PN} (R)$, Eq.~(\ref{eq7.5}), 
is rather large and positive indicates that the $\nu$-dependent 
nonlinear gravitational effects lead, for comparable 
masses $(\nu \sim \frac{1}{4}$), to a last stable 
(circular) orbit (LSO) which has a higher frequency and 
a larger binding energy than what a naive scaling from 
the test-particle limit $(\nu \to 0)$ would suggest. 
Actually, the PN-expanded form (\ref{eq7.5}) of $A_{\rm 3PN} (R)$ 
does not seem to be a good representation of the (unknown) exact 
function $A_{\rm EOB} (R)$ when the (Schwarzschild-like) relative 
coordinate $R$ becomes smaller than about $6 GM / c^2$ (which is 
the radius of the LSO in the test-mass limit). 
By continuity with the test-mass case, one a priori
expects that $A_{\rm 3PN}(R)$ always exhibits a simple zero 
defining an EOB ``effective horizon'' that is smoothly connected
to the Schwarzschild event horizon at $R= 2GM/c^2$ when $\nu\to 0$. 
However, the large value of the $a_4$ coefficient does actually 
prevent $A_{\rm 3PN}$ to have this property
when $\nu$ is too large, and in particular when $\nu=1/4$. It was therefore suggested~\cite{DJS2000} to further resum\footnote{The PN-expanded EOB 
building blocks $A_{3{\rm PN}} (R) , B_{3{\rm PN}} (R) , \ldots$ already 
represent a \textit{resummation} of the PN dynamics 
in the sense that they have ``condensed'' the many 
terms of the original PN-expanded Hamiltonian within 
a very concise format. But one should not refrain to 
further resum the EOB building blocks themselves, if 
this is physically motivated.} $A_{\rm 3PN} (R)$ by 
replacing it by a suitable Pad\'e $(P)$ approximant. 
For instance, the replacement of $A_{\rm 3PN} (R)$ by\footnote{We recall that the coefficients $n_1$ and $(d_1,d_2,d_3)$ of the $(1,3)$ Pad\'e approximant $P_3^1 [A_{\rm 3PN} (u)]$ are determined 
by the condition that the first four terms of the Taylor expansion 
of $A_3^1$ in powers of $u=GM/(c^2R)$ coincide with $A_{\rm 3PN}$.}
\begin{equation}
\label{eq7.7}
A_3^1 (R) \equiv P_3^1 [A_{\rm 3PN} (R)] = \frac{1+n_1 u}{1+d_1 u + d_2 u^2 + d_3 u^3}
\end{equation}
ensures that the $\nu = \frac{1}{4}$ case is smoothly 
connected with the $\nu = 0$ limit.

\medskip

The same kind of $\nu$-continuity argument, discussed so far for the $A$ function, needs to be applied also to the $D_{\rm 3PN}(R)$ function defined in Eq.~(\ref{eq:D}). 
A straightforward way to ensure that the $D$ function stays positive when $R$ decreases (since it is $D=1$ when $\nu\to 0$) is to replace $D_{\rm 3PN}(R)$ by $D^0_3(R)\equiv P^0_3\left[D_{\rm 3PN}(R)\right]$, where $P^0_3$ indicates the $(0,3)$ Pad\'e approximant and explicitly reads
\begin{equation}
\label{eq25}
D^0_3(R)=\frac{1}{1+6\nu u^2  -2(3\nu-26)\nu u^3}.
\end{equation}

\section{EOB description of radiation reaction and of the emitted waveform during inspiral} 

In the previous Section we have described how the EOB method 
encodes the conservative part of the relative orbital dynamics 
into the dynamics of an 'effective' particle. Let us now 
briefly discuss how to complete the EOB dynamics by defining 
some \textit{resummed} expressions describing radiation reaction effects, and the corresponding waveform emitted at infinity. 
One is interested in circularized binaries, which have lost 
their initial eccentricity under the influence of radiation 
reaction. For such systems, it is enough 
(in first approximation~\cite{Buonanno2000}; see, however, the recent results of Bini and Damour \cite{Bini2012}) to include a radiation reaction force in the $P_{\varphi}$ equation 
of motion only. More precisely, we are using phase space variables $R, P_R , \varphi , P_{\varphi}$ associated to polar coordinates (in the equatorial plane 
$\theta = \frac{\pi}{2}$). Actually it is convenient to replace the radial momentum $P_R$ by the momentum conjugate to the `tortoise' radial coordinate $R_* = \int dR (B/A)^{1/2}$, \textit{i.e.} $P_{R_*} = (A/B)^{1/2} \,P_R$. The real EOB Hamiltonian is obtained by first solving Eq.~(\ref{eqn9}) to get $H_{\rm real}^{\rm total} = \sqrt s$ in terms of ${\mathcal E}_{\rm eff}$, and then by solving the effective Hamilton-Jacobi equation to get ${\mathcal E}_{\rm eff}$ in terms of the effective phase space coordinates ${\bm q}_{\rm eff}$ and ${\bm p}_{\rm eff}$. The result is given by two nested square roots (we henceforth set $c=1$):
\begin{equation}
\label{eqn10}
\hat H_{\rm EOB} (r,p_{r_*} , \varphi) = \frac{H_{\rm EOB}^{\rm real}}{\mu} 
= \frac{1}{\nu}\sqrt{1 + 2 \nu \, (\hat H_{\rm eff} - 1)} \, ,
\end{equation}
where
\begin{equation}
\label{eqn11}
\hat H_{\rm eff} = \sqrt{p_{r_*}^2 + A(r) \left( 1 + \frac{p_{\varphi}^2}{r^2} + z_3 \, \frac{p_{r_*}^4}{r^2} \right)} \, ,
\end{equation}
with $z_3 = 2\nu \, (4-3\nu)$. Here, we are using suitably rescaled
dimensionless (effective) variables: 
$r = R/GM$, $p_{r_*} = P_{R_*} / \mu$, $p_{\varphi} = P_{\varphi} / \mu \, 
GM$, as well as a rescaled time $t = T/GM$. This leads to equations 
of motion for $(r,\varphi,p_{r_*},p_{\varphi})$ of the form
\begin{eqnarray}
\label{eqn12}
\frac{d \varphi}{dt}     & =&   \frac{\partial \, \hat H_{\rm EOB}}{\partial \, p_{\varphi}}\equiv\Omega \, ,\\
\label{eqn12a}
\frac{dr}{dt}           & =  & \left( \frac{A}{B} \right)^{1/2} \,
\frac{\partial \, \hat H_{\rm EOB}}{\partial \, p_{r_*}} \, ,\\
\label{eqn12b}
\frac{d p_{\varphi}}{dt} & = &  \hat{\mathcal F}_{\varphi} \,,\\ 
\label{eqn12c}
\frac{d p_{r_*}}{dt}    & = &- \left( \frac{A}{B} \right)^{1/2} \, \frac{\partial \, \hat H_{\rm EOB}}{\partial \, r} \, ,
\end{eqnarray}
which explicitly read
\begin{eqnarray}
\label{eob:1}
\frac{d\varphi}{dt}     &= &\frac{A p_\varphi}{\nu r^2\hat{H}\hat{H}_{\rm eff}} \equiv \Omega\ , \\
\label{eob:2}
\frac{dr}{dt}           &=& \left(\frac{A}{B}\right)^{1/2}\frac{1}{\nu\hat{H}\hat{H}_{\rm eff}}\left(p_{r_*}+z_3\frac{2A}{r^2}p_{r_*}^3\right) \ , \\
\label{eob:3}
\frac{dp_{\varphi}}{dt} &=& \hat{\cal F}_{\varphi} \ , \\
\frac{dp_{r_*}}{dt}     &= &-\left(\frac{A}{B}\right)^{1/2}\frac{1}{2\nu\hat{H}\hat{H}_{\rm eff}} \nonumber \\
\label{eob:4}
&&\left\{A'+\frac{p_\varphi^2}{r^2}\left(A'-\frac{2A}{r}\right)+z_3\left(\frac{A'}{r^2}-\frac{2A}{r^3}\right)p_{r_*}^4 \right\} \ ,
\end{eqnarray}
where $A'=dA/dr$. As explained above the EOB metric function $A(r)$ is defined by 
Pad\'e resumming the Taylor-expanded result (\ref{eqn4}) obtained  from the matching between the real and effective energy levels (as we were mentioning, one uses a similar Pad\'e resumming for $D(r) \equiv A(r) \, B(r)$). One similarly needs to resum 
${\hat{\cal F}}_{\varphi}$, i.e., the $\varphi$ component of the radiation reaction which has been introduced on the r.h.s. of Eq.~(\ref{eqn12b}). 

\medskip

Several methods have been tried during the development of the EOB formalism to resum the radiation reaction $\widehat{\mathcal F}_{\varphi}$ (starting from the high-order PN-expanded results that have been obtained in the literature; see references in the Introduction above). Here, we shall briefly explain the new, \textit{parameter-free} resummation technique for the multipolar waveform (and thus for the energy flux) introduced
in Ref.~\cite{Damour2007,Damour2008} and perfected in~\cite{Damour2009}. 
To be precise, the new results discussed in Ref.~\cite{Damour2009} are twofold: on the one hand, that work generalized the $\ell=m=2$  \textit{ resummed factorized waveform} of~\cite{Damour2007,Damour2008} to higher multipoles by using the most accurate currently known PN-expanded results~\cite{Kidder2008,Berti2007,BFIS2008}
as well as the higher PN terms which are known in the test-mass
limit~\cite{Tagoshi1994,Tanaka1996}; on the other hand,
it introduced a \textit{further resummation procedure} which consists in 
considering a new theoretical quantity, denoted as $\rho_{\ell m}(x)$, 
which enters the $(\ell,m)$ waveform (together with other building 
blocks, see below) only through its $\ell$-th power: $h_{\ell m}\propto
\left(\rho_{\ell m}(x)\right)^{\ell}$. Here, and below, $x$ denotes the
invariant PN-ordering parameter given during inspiral by $x\equiv (GM\Omega/c^3)^{2/3}$.

\medskip

The main novelty introduced by Refs.~\cite{Damour2007,Damour2008,Damour2009} is 
to write the $(\ell,m)$ multipolar waveform emitted by a 
circular nonspinning compact binary as the \textit{product} of several 
factors, namely
\begin{equation}
\label{eq:hlm}
h_{\ell m}^{(\epsilon)}=\frac{GM\nu}{c^2 R} n_{\ell m}^{(\epsilon)} c_{\l+\epsilon}(\nu)
x^{(\ell+\epsilon)/2}Y^{\ell-\epsilon,-m}\left(\frac{\pi}{2},\Phi\right)
\hat{S}_{\rm  eff}^{(\epsilon)}T_{\ell m} e^{{\rm i}\delta_{\ell m}} \rho_{\ell m}^\ell.
\end{equation}
Here $\epsilon$ denotes the parity of $\ell+m$ ($\epsilon=\pi(\ell+m)$), i.e.
$\epsilon=0$ for ``even-parity'' (mass-generated) multipoles ($\ell+m$ even), and
$\epsilon=1$ for ``odd-parity'' (current-generated) ones ($\ell+m$ odd); $n_{\ell m}^{(\epsilon)}$ and $c_{\l+\epsilon}(\nu)$ are numerical coefficients; $\hat{S}^{(\epsilon)}_{\rm eff}$ is a $\mu$-normalized effective source (whose definition comes from the EOB formalism); $T_{\ell m}$ is a resummed version~\cite{Damour2007,Damour2008} of an infinite number of
``leading logarithms'' entering the \textit{tail effects}~\cite{Blanchet1992,Blanchet1998};
$\delta_{\ell m}$ is a supplementary phase (which corrects the phase effects not
included in the \textit{complex} tail factor $T_{\ell m}$), and, finally,
$\left(\rho_{\ell m}\right)^\ell$ denotes the $\ell$-th power of the quantity
$\rho_{\ell m}$ which is the new building block introduced
in~\cite{Damour2009}. Note that in previous
papers~\cite{Damour2007,Damour2008}  the quantity 
$\left(\rho_{\ell m}\right)^\ell$ was denoted as $f_{\ell m}$
and we will often use this notation below.
Before introducing explicitly the various elements entering the 
waveform (\ref{eq:hlm}) it is convenient to decompose $h_{\ell m}$ as 
\begin{equation}
\label{hlm_expanded}
h_{\ell m}^{(\epsilon)} = h_{\ell m}^{(N,\epsilon)} \hat{h}_{\ell m}^{(\epsilon)},
\end{equation}
where $h_{\ell m}^{(N,\epsilon)}$ is the Newtonian contribution (\textit{i.e.} the product of the first five factors in Eq.~(\ref{eq:hlm})) and 
\begin{equation}
\hat{h}_{\ell m}^{(\epsilon)}\equiv \hat{S}_{\rm eff}^{(\epsilon)} T_{\ell m}e^{\rm i\delta_{\ell m}}f_{\ell m}
\end{equation}
represents a resummed version of all the PN corrections. The PN correcting factor 
$\hat{h}_{\ell m}^{(\epsilon)}$, as well as all its building blocks, has the structure
$\hat{h}^{(\epsilon)}_{\ell m}=1+{\cal O}(x)$.

\medskip

The reader will find in Ref.~\cite{Damour2009} the definitions of the quantities entering the ``Newtonian'' waveform $h_{\ell m}^{(N,\epsilon)}$, as well as the precise definition of the effective source factor $\widehat S_{\rm eff}^{(\epsilon)}$, which constitutes the first factor in the PN-correcting factor $\widehat h_{\ell m}^{(\epsilon)}$. Let us only note here that the definition of $\widehat S_{\rm eff}^{(\epsilon)}$ makes use of EOB-defined quantities. For instance, for even-parity waves $(\epsilon=0)$ $\widehat S_{\rm eff}^{(0)}$ is defined as the $\mu$-scaled \textit{effective} energy ${\mathcal E}_{\rm eff} / \mu c^2$. [We use the ``$J$-factorization'' definition of $\widehat S_{\rm eff}^{(\epsilon)}$ when $\epsilon = 1$, \textit{i.e.} for odd parity waves.]

\medskip

The second building block in the factorized decomposition is the ``tail
factor'' $T_{\ell m}$ (introduced in Refs.~\cite{Damour2007,Damour2008}).
As mentioned above, $T_{\ell m}$ is a resummed version of an infinite number 
of ``leading logarithms'' entering the transfer function between the 
near-zone multipolar wave and the far-zone one, 
due to \textit{tail effects} linked to its propagation in a Schwarzschild 
background of mass $M_{\rm ADM}=H^{\rm real}_{\rm EOB}$. 
Its explicit expression reads
\begin{equation}
\label{eq:tail_factor}
T_{\ell m} = \frac{\Gamma(\ell+1-2{\rm i}\hat{\hat{k}})}{\Gamma(\ell
  +1)}e^{\pi\hat{\hat{k}}}e^{2{\rm i}\hat{\hat{k}}\log(2 k r_0)} ,
\end{equation}
where $r_0=2GM/\sqrt e$ and $\hat{\hat{k}}\equiv G H^{\rm real}_{\rm EOB} m\Omega$
and $k\equiv m\Omega$.
Note that  $\hat{\hat{k}}$ differs from $k$ by a rescaling involving 
the \textit{real} (rather than the \textit{effective}) 
EOB Hamiltonian, computed at this stage along the sequence of
circular orbits.

\medskip

The tail factor $T_{\ell m}$ is a complex number which already takes into 
account some of the dephasing of the partial waves as they propagate
out from the near zone to infinity. However, as the tail factor only takes 
into account the leading logarithms, one needs to correct it by a complementary 
dephasing term, $e^{{\rm i}\delta_{\ell m}}$,  linked to subleading logarithms and other effects.
This subleading phase correction can be computed as being the phase
$\delta_{\ell m}$ of the
complex ratio between the PN-expanded $\hat{h}_{\ell m}^{(\epsilon)}$ and the 
above defined source and tail factors. In the comparable-mass case
($\nu\neq0$), the 3PN $\delta_{22}$ phase correction to the leading quadrupolar
wave was originally computed in Ref.~\cite{Damour2008} (see also
Ref.~\cite{Damour2007} for the $\nu=0$ limit). Full results for
the subleading partial waves to the highest  
possible PN-accuracy by starting from the currently known 
3PN-accurate $\nu$-dependent waveform~\cite{BFIS2008}
have been obtained in~\cite{Damour2009}. For higher-order test-mass $(\nu \to 0)$ contributions, see \cite{Fujita2010,Fujita2012}. For extensions of the (non spinning) factorized waveform of \cite{Damour2009} see \cite{YPan2011,Pan2011,Taracchini2012}.

\medskip

The last factor in the multiplicative decomposition
of the multipolar waveform can be computed 
as being the modulus $f_{\ell m}$ of the complex ratio between 
the PN-expanded $\hat{h}_{\ell m}^{(\epsilon)}$  and the 
above defined source and tail factors.
In the comparable mass case
($\nu\neq0$), the $f_{22}$ modulus correction to the leading quadrupolar
wave was computed in Ref.~\cite{Damour2008} (see also
Ref.~\cite{Damour2007} for the $\nu=0$ limit). 
For the  subleading partial waves, Ref.~\cite{Damour2009}
explicitly computed the other $f_{\ell m}$'s to the highest 
possible PN-accuracy by starting from the currently known 
3PN-accurate $\nu$-dependent waveform~\cite{BFIS2008}.
In addition, as originally proposed in Ref.~\cite{Damour2008}, 
to reach greater accuracy the $f_{\ell m}(x;\nu)$'s extracted from
the 3PN-accurate $\nu\neq 0$ results  are completed by adding 
higher order contributions coming from the 
$\nu=0$ results~\cite{Tagoshi1994,Tanaka1996}.
In the particular $f_{22}$ case discussed 
in~\cite{Damour2008}, this amounted to adding 4PN and 5PN $\nu=0$
terms. This ``hybridization'' procedure was then systematically 
pursued for all the other multipoles, using the 5.5PN accurate 
calculation of the multipolar decomposition of the gravitational 
wave energy flux of Refs.~\cite{Tagoshi1994,Tanaka1996}.

\medskip

The decomposition of the total PN-correction factor $\hat{h}_{\ell m}^{(\epsilon)}$
into several factors is in itself a resummation procedure which already
improves the convergence of the PN series one has to deal with:
indeed, one can see that the coefficients entering increasing powers of $x$ in the PN expansion of the $f_{\ell m}$'s tend to be systematically smaller than the coefficients appearing in the usual PN expansion of $\hat{h}_{\ell m}^{(\epsilon)}$. The reason for this
is essentially twofold: (i) the factorization of $T_{\ell m}$ has absorbed powers 
of $m\pi$ which contributed to make large coefficients in $\hat{h}_{\ell m}^{(\epsilon)}$,
and (ii) the factorization of either $\hat{H}_{\rm eff}$ or $\hat{j}$ has (in the $\nu=0$ case)  removed the presence of an inverse square-root singularity located at $x=1/3$ 
which caused the coefficient of $x^n$ in any PN-expanded quantity to grow
as $3^{n}$ as $n\to\infty$.

\medskip

To further improve the convergence of the waveform several resummations of the factor $f_{\ell m} (x) = 1 + c_1^{\ell m} x + c_2^{\ell m} x^2 + \ldots$ have been suggested. First, Refs.~\cite{Damour2007,Damour2008} proposed to further resum the $f_{22}(x)$ function via a Pad\'e (3,2) approximant, $P^3_{2}\{f_{22}(x;\nu)\}$, so as to improve its behavior in the strong-field-fast-motion regime. Such a resummation gave an excellent
agreement with numerically computed waveforms, near the end of the inspiral 
and during the beginning of the plunge, for different mass ratios~\cite{Damour2007,DNNPR2008,DNHHB2008}. As we were mentioning above, a new route for resumming $f_{\ell m}$ was explored in Ref.~\cite{Damour2009}. It is 
based on replacing $f_{\ell m}$ by its $\ell$-th root, say
\begin{equation}
\label{eq:lth_root}
\rho_{\ell m}(x;\nu) = [f_{\ell m}(x;\nu)]^{1/\ell}.
\end{equation}
The basic motivation for replacing $f_{\ell m}$ by $\rho_{\ell m}$ 
is the following: the leading ``Newtonian-level'' contribution 
to the waveform $h^{(\epsilon)}_{\ell m}$ contains a  factor 
$\omega^\ell r_{\rm harm}^\ell v^\epsilon$ where $r_{\rm harm}$  is the
harmonic radial coordinate used in the MPM 
formalism~\cite{Blanchet1989,DamourIyer1991}.
When computing the PN expansion of this factor one has to insert 
the PN expansion of the (dimensionless) harmonic radial 
coordinate $r_{\rm harm}$, $ r_{\rm harm} = x^{-1}(1+c_1 x+{\cal O }(x^2))$,
as a function of the gauge-independent
frequency parameter $x$. 
The PN re-expansion of $[r_{\rm harm}(x)]^\ell$ then generates terms of the 
type $x^{-\ell}(1 +\ell c_1 x+....)$. 
This is one (though not the only one) of the origins of 
1PN corrections in $h_{\ell m}$ and $f_{\ell m}$ 
whose coefficients grow linearly with $\ell$.
The study of~\cite{Damour2009} has pointed out that
these $\ell$-growing terms are problematic for
the accuracy of the PN-expansions. 
The replacement of $f_{\ell m}$ by $\rho_{\ell m}$ 
is a cure for this problem.

\medskip

Several studies, both in the test-mass limit, $\nu \to 0$ (see Fig.~1 in \cite{Damour2009}) and in the comparable-mass case (see notably Fig.~4 in \cite{DamourNagar2009}), have shown that the resummed factorized (inspiral) EOB waveforms defined above provided remarkably accurate analytical approximations to the ``exact'' inspiral waveforms computed by numerical simulations. These resummed multipolar EOB waveforms are much closer (especially during late inspiral) to the exact ones than the standard PN-expanded waveforms given by Eq.~(\ref{hlm_expanded}) with a PN-correction factor of the usual ``Taylor-expanded'' form
$$
\widehat h_{\ell m}^{(\epsilon) {\rm PN}} = 1 + c_1^{\ell m} x + c_{3/2}^{\ell m} x^{3/2} + c_2^{\ell m} x^2 + \ldots
$$
See Fig.~1 in \cite{Damour2009}, and slide 29 in my (June 2012) Prague presentation.

\medskip

Finally, one uses the newly resummed multipolar waveforms~(\ref{eq:hlm})
to define a resummation of the \textit{radiation reaction force} 
$\cal F_{\varphi}$  as
\begin{equation}
\label{eq:RR_new}
{\cal F}_{\varphi} \equiv -\frac{1}{\Omega} F^{(\ell_{\rm max})},
\end{equation}
where the (instantaneous, circular) GW flux $F^{(\ell_{\rm max})}$ is defined 
as 
\begin{equation}
\label{eq:flux_1}
 F^{(\ell_{\rm max})}\equiv\frac{2}{16\pi G}\sum_{\ell =2}^{\ell_{\rm max}}\sum_{m=1}^{\ell}(m\Omega)^2|R h_{\ell m}|^2.
\end{equation}

\medskip

Summarizing: Eqs.~(\ref{eq:hlm}) and (\ref{eq:RR_new}), (\ref{eq:flux_1}) define resummed EOB versions of the waveform $h_{\ell m}$, and of the radiation reaction $\widehat{\mathcal F}_{\varphi}$, during inspiral. A crucial point is that these resummed expressions are \textit{parameter-free}. Given some current approximation to the conservative EOB dynamics (\textit{i.e.} some expressions for the $A,D,Q$ potentials) they \textit{complete} the EOB formalism by giving explicit predictions for the radiation reaction (thereby completing the dynamics, see Eqs.~(\ref{eqn12})--(\ref{eqn12c})), and for the emitted inspiral waveform.

\section{EOB description of the merger of binary black holes and of the ringdown of the final black hole}

Up to now we have reviewed how the EOB formalism, starting only from \textit{analytical} information obtained from PN theory, and adding extra resummation requirements (both for the EOB conservative potentials $A$, Eq.~(\ref{eq7.7}), and $D$, Eq.~(\ref{eq25}), and for the waveform, Eq.~(\ref{eq:hlm}), and its associated radiation reaction force, Eqs.~(\ref{eq:RR_new}), (\ref{eq:flux_1})) makes specific predictions, both for the motion and the radiation of binary black holes. The analytical calculations underlying such an EOB description are essentially based on skeletonizing the two black holes as two, sufficiently separated point masses, and therefore seem unable to describe the merger of the two black holes, and the subsequent ringdown of the final, single black hole formed during the merger. However, as early as 2000 \cite{Buonanno2000}, the EOB formalism went one step further and proposed a specific strategy for describing the \textit{complete} waveform emitted during the entire coalescence process, covering inspiral, merger and ringdown. This EOB proposal is somewhat crude. However, the predictions it has made (years before NR simulations could accurately describe the late inspiral and merger of binary black holes)  have been broadly confirmed by subsequent NR simulations. [See the Introduction for a list of EOB predictions.] The original EOB proposal (which was motivated partly by the closeness between the 2PN-accurate effective metric $g_{\mu\nu}^{\rm eff}$ \cite{Buonanno1999} and the Schwarzschild metric, and by the results of Refs.~\cite{Davis1972} and \cite{Price1994}) consists of:

\medskip

(i) defining, within EOB theory, the instant of (effective) ``\textit{merger}'' of the two black holes as the (dynamical) EOB time $t_m$ where the orbital frequency $\Omega (t)$ reaches its \textit{maximum};

\medskip

(ii) describing (for $t \leq t_m$) the inspiral-plus-plunge (or simply \textit{insplunge}) waveform, $h^{\rm insplunge} (t)$, by using the inspiral EOB dynamics and waveform reviewed in the previous Section; and

\medskip

(iii) describing (for $t \geq t_m$) the merger-plus-ringdown waveform as a superposition of several quasi-normal-mode (QNM) complex frequencies of a final Kerr black hole (of mass $M_f$ and spin parameter $a_f$, self-consistency estimated within the EOB formalism), say 
\begin{equation}
\label{eqn17}
\left( \frac{R c^2}{GM} \right) h_{\ell m}^{\rm ringdown} (t) = \sum_N C_N^+ \, e^{-\sigma_N^+ (t-t_m)} \, ,
\end{equation}
with $\sigma_N^+ = \alpha_N + i \, \omega_N$, and where the label $N$ refers
to indices $(\ell , \ell' , m , n)$, with $(\ell , m)$ being the 
Schwarzschild-background multipolarity of the considered (metric) waveform 
$h_{\ell m}$, with $n=0,1,2\ldots$ being the `overtone number' of the 
considered Kerr-background Quasi-Normal-Mode, and $\ell'$ the degree of 
its associated spheroidal harmonics $S_{\ell ' m} (a \sigma , \theta)$;

\medskip

(iv) determining the excitation coefficients $C_N^+$ of the QNM's in Eq.~(\ref{eqn17}) by using a simplified representation of the transition between plunge and ring-down obtained by smoothly \textit{matching} (following Ref.~\cite{Damour2007}), on a $(2p+1)$-toothed ``comb'' $(t_m - p\delta , \ldots , t_m - \delta , t_m , t_m + \delta , \ldots, t_m + p\delta)$ centered around the merger (and matching) time $t_m$, the inspiral-plus-plunge waveform to the above ring-down waveform.

\medskip

Finally, one defines a complete, quasi-analytical EOB waveform 
(covering the full process from inspiral to ring-down) as:
\begin{equation}
\label{eqn18}
h_{\ell m}^{\rm EOB} (t) = \theta (t_m - t) \, h_{\ell m}^{\rm insplunge} (t) + \theta (t-t_m) \, h_{\ell m}^{\rm ringdown} (t) \, ,
\end{equation}
where $\theta (t)$ denotes Heaviside's step function. The final result is a waveform that essentially depends only on the choice of a resummed EOB $A(u)$ potential, and, less importantly, on the choice of resummation of the main waveform amplitude factor $f_{22} = (\rho_{22})^2$.

\medskip

We have emphasized here that the EOB formalism is able, in principle, starting only from the best currently known analytical information, to predict the full waveform emitted by coalescing binary black holes. The early comparisons between 3PN-accurate EOB predicted waveforms\footnote{The new, resummed EOB waveform discussed above was not available at the time, so that these comparisons employed the coarser ``Newtonian-level'' EOB waveform $h_{22}^{(N,\epsilon)} (x)$.} and NR-computed waveforms showed a satisfactory agreement between the two, within the (then relatively large) NR uncertainties \cite{Buonanno2007,Pan2008}. Moreover, as we shall discuss below, it has been recently shown that the currently known Pad\'e-resummed 3PN-accurate $A(u)$ potential is able, as is, to describe with remarkable accuracy several aspects of the dynamics of coalescing binary black holes, \cite{LeTiec2011,Damour2012}.

\medskip

On the other hand, when NR started delivering high-accuracy waveforms, it became clear that the 3PN-level analytical knowledge incorporated in EOB theory was not accurate enough for providing waveforms agreeing with NR ones within the high-accuracy needed for detection, and data analysis of upcoming GW signals. [See, \textit{e.g.}, the discussion in Section~II of Ref.~\cite{Pan2011}.] At that point, one made use of the \textit{natural flexibility} of the EOB formalism. Indeed, as already emphasized in early EOB work \cite{TDamour2001,Damour2002}, we know from the analytical point of view that there are (yet uncalculated) further terms in the $u$-expansions of the EOB potentials $A(u), D(u), \ldots$ (and in the $x$-expansion of the waveform), so that these terms can be introduced either as ``free parameter(s) in constructing a bank of templates, and [one should] wait until'' GW observations determine their value(s) \cite{TDamour2001}, or as ``\textit{fitting parameters} and adjusted so as to reproduce other information one has about the exact results'' (to quote Ref.~\cite{Damour2002}). For instance, modulo logarithmic corrections that will be further discussed below, the Taylor expansion in powers of $u$ of the main EOB potential $A(u)$ reads
$$
A^{\rm Taylor} (u;\nu) = 1-2u + \widetilde a_3 (\nu) u^3 + \widetilde a_4 (\nu) u^4 + \widetilde a_5 (\nu) u^5 + \widetilde a_6 (\nu) u^6 + \ldots
$$
where the 2PN and 3PN coefficients $\widetilde a_3 (\nu) = 2\nu$ and $\widetilde a_4 (\nu) = a_4 \nu$ are known, but where the 4PN, 5PN,$\ldots$ coefficients, $\widetilde a_5 (\nu) , \widetilde a_6 (\nu), \ldots$ have not yet been calculated (see, however, below). A first attempt was made in \cite{Damour2002} to use numerical data (on circular orbits of corotating black holes) to fit for the value of a (single, effective) 4PN parameter of the simple form $\widetilde a_5 (\nu) = a_5 \nu$ entering a Pad\'e-resummed 4PN-level $A$ potential, \textit{i.e.}
\begin{equation}
\label{A_4PN}
A^1_4(u;a_5,\nu) = P^{1}_{4}\left[A_{\rm 3PN}(u) + \nu a_5 u^5 \right] \, .
\end{equation}
This strategy was pursued in Ref.~\cite{ABuonanno2007,Damour2008} and many subsequent works. It was pointed out in Ref.~\cite{DamourNagar2009} that the introduction of a further 5PN coefficient $\widetilde a_6 (\nu) = a_6 \nu$, entering a Pad\'e-resummed 5PN-level $A$ potential, \textit{i.e.}
\begin{equation}
\label{A_5PN}
A^1_5(u;a_5,a_6,\nu) = P^{1}_{5}\left[A_{\rm 3PN}(u) + \nu a_5 u^5 + \nu a_6 u^6\right] \, ,
\end{equation}
helped in having a closer agreement with accurate NR waveforms.

\medskip

In addition, Refs.~\cite{Damour2007,Damour2008} introduced another type of flexibility parameters of the EOB formalism: the non quasi-circular (NQC) parameters accounting for  uncalculated modifications of the quasi-circular inspiral waveform presented above, linked to deviations from  an adiabatic quasi-circular motion. These NQC parameters are of various types, and subsequent works \cite{DNNPR2008,DNHHB2008,DamourNagar2009,Buonanno2009,Bernuzzi2011,Barausse:2011kb,Pan2011} have explored several ways of introducing them. They enter the EOB waveform in two separate ways. First, through an explicit, additional complex factor multiplying $h_{\ell m}$, \textit{e.g.}
$$
f_{\ell m}^{\rm NQC} = (1+a_1^{\ell m} n_1 + a_2^{\ell m} n_2) \exp [i(a_3^{\ell m} n_3 + a_4^{\ell m} n_4)]
$$
where the $n_i$'s are dynamical functions that vanish in the quasi-circular limit (with $n_1 , n_2$ being time-even, and $n_3 , n_4$ time-odd). For instance, one usually takes $n_1 = (p_{r_*} / r\Omega)^2$. Second, through the (discrete) choice of the argument used during the plunge to replace the variable $x$ of the quasi-circular inspiral argument: \textit{e.g.} either $x_{\Omega} \equiv (GM \Omega)^{2/3}$, or (following \cite{Damour2006}) $x_{\varphi} \equiv v_{\varphi}^2 = (r_{\omega} \Omega)^2$ where $v_{\varphi} \equiv \Omega \, r_{\omega}$, and $r_{\omega}\equiv r[\psi(r,p_\varphi)]^{1/3}$ is a 
modified EOB radius, with $\psi$ being defined as 
\begin{equation}
\psi(r,p_\varphi)=\frac{2}{r^2}\left(\frac{dA(r)}{dr}\right)^{-1} \left[1+2\nu\left(\sqrt{A(r)\left(1+\frac{p_\varphi^2}{r^2}\right)  }-1\right)\right] \, .
\end{equation}
For a given value of the symmetric mass ratio, and given values of the $A$-flexibility parameters $\widetilde a_5 (\nu) , \widetilde a_6 (\nu)$ one can determine the values of the NQC parameters $a_i^{\ell m}$'s from accurate NR simulations of binary black hole coalescence (with mass ratio $\nu$) by imposing, say, that the complex EOB waveform $h_{\ell m}^{\rm EOB} (t^{\rm EOB} ; \widetilde a_5 , \widetilde a_6; a_i^{\ell m})$ \textit{osculates} the corresponding NR one $h_{\ell m}^{\rm NR} (t^{\rm NR})$ at their respective instants of ``merger'', where $t_{\rm merger}^{\rm EOB} \equiv t_m^{\rm EOB}$ was defined above (maximum of $\Omega^{\rm EOB} (t)$), while $t_{\rm merger}^{\rm NR}$ is defined, say, as the (retarded) NR time where the modulus $\vert h_{22}^{\rm NR} (t) \vert$ of the quadrupolar waveform reaches its maximum. The order of osculation that one requires between $h_{\ell m}^{\rm EOB} (t)$ and $h_{\ell m}^{\rm NR} (t)$ (or, separately, between their moduli and their phases or frequencies) depends on the number of NQC parameters $a_i^{\ell m}$. For instance, $a_1^{\ell m}$ and $a_2^{\ell m}$ affect only the modulus of $h_{\ell m}^{\rm EOB}$ and allow one to match both $\vert h_{\ell m}^{\rm EOB} \vert$ and its first time derivative, at merger, to their NR counterparts, while $a_3^{\ell m}, a_4^{\ell m}$ affect only the phase of the EOB waveform, and allow one to match the GW frequency $\omega_{\ell m}^{\rm EOB} (t)$ and its first time derivative, at merger, to their NR counterparts. The above EOB/NR matching scheme has been developed and declined in various versions in Refs.~\cite{DNNPR2008,DNHHB2008,DamourNagar2009,Buonanno2009,Bernuzzi2011,SBernuzzi2011,Barausse:2011kb,Pan2011,DNB2012}. One has also extracted the needed matching data from accurate NR simulations, and provided explicit, analytical $\nu$-dependent fitting formulas for them \cite{DamourNagar2009,Pan2011,DNB2012}.

\medskip

Having so ``calibrated'' the values of the NQC parameters by extracting non-perturbative information from a sample of NR simulations, one can then, for any choice of the $A$-flexibility parameters, compute a full EOB waveform (from early inspiral to late ringdown). The comparison of the latter NQC-completed EOB waveform to the results of NR simulations is discussed in the next Section.

\section{EOB vs NR}

There have been several different types of comparison between EOB and NR. For instance, the early work \cite{Buonanno2007} pioneered the comparison between a purely analytical EOB waveform (uncalibrated to any NR information) and a NR waveform, while the early work \cite{DamourNagar2007} compared the predictions for the final spin of a coalescing black hole binary made by EOB, \textit{completed} by the knowledge of the energy and angular momentum lost during ringdown by an extreme mass ratio binary (computed by the test-mass NR code of \cite{Nagar2007}), to comparable-mass NR simulations \cite{Gonzalez2007}. Since then, many other EOB/NR comparisons have been performed, both in the comparable-mass case \cite{Pan2008,ABuonanno2007,Damour2008,DNNPR2008,DNHHB2008,DamourNagar2009,Buonanno2009}, and in the small-mass-ratio case \cite{Damour2007,Yunes2010,Yunes2011,Bernuzzi2011,Barausse:2011kb}. Note in this respect that the numerical simulations of the GW emission by extreme mass-ratio binaries have provided (and still provide) a very useful ``laboratory'' for learning about the motion and radiation of binary systems, and their description within the EOB formalism.

\medskip

Here we shall discuss only two recent examples of EOB/NR comparisons, which illustrate different facets of this comparison.

\subsection{EOB[NR] waveforms vs NR ones}

We explained above how one could complete the EOB formalism by calibrating some of the natural EOB flexibility parameters against NR data. First, for any given mass ratio $\nu$ and any given values of the $A$-flexibility parameters $\widetilde a_5 (\nu) , \widetilde a_6 (\nu)$, one can use NR data to uniquely determine the NQC flexibility parameters $a_i$'s. In other words, we have (for a given $\nu$)
$$
a_i = a_i [{\rm NR \, data} ; a_5 , a_6] \, ,
$$
where we defined $a_5$ and $a_6$ so that $\widetilde a_5 (\nu) = a_5 \nu , \widetilde a_6 (\nu) = a_6 \nu$. [We allow for some residual $\nu$-dependence in $a_5$ and $a_6$.] Inserting these values in the (analytical) EOB waveform then defines an NR-completed EOB waveform which still depends on the two unknown flexibility parameters $a_5$ and $a_6$.

\medskip

In Ref.~\cite{DamourNagar2009} the $(a_5,a_6)$-dependent 
predictions made by such a NR-completed EOB formalism were compared to 
the high-accuracy waveform from an equal-mass binary black hole ($\nu=1/4$) 
computed by the Caltech-Cornell-CITA group~\cite{Scheel2009},
(and then made available on the web). It was found that there is a strong degeneracy between $a_5$ and $a_6$ in the sense that there is an excellent EOB-NR agreement 
for an extended region in the $(a_5,a_6)$-plane. More precisely, the phase difference between the EOB (metric) waveform and the Caltech-Cornell-CITA one, considered
between GW frequencies $M\omega_{\rm L}=0.047$ and 
$M\omega_{\rm R}=0.31$ (i.e., the last 16 GW cycles before merger),
stays smaller than 0.02 radians within a long and thin 
banana-like region in the $(a_5,a_6)$-plane. This ``good region'' 
approximately extends between the points 
$(a_5,a_6)=(0,-20)$ and $(a_5,a_6)=(-36,+520)$. 
As an example (which actually lies on the boundary of the 
``good region''), we shall consider here (following Ref.~\cite{DamourNagar2011}) the specific values $a_5=0, a_6=-20$ 
(to which correspond, when $\nu=1/4$, $a_1 = -0.036347, a_2=1.2468$). [Ref.~\cite{DamourNagar2009} did not make use of the NQC phase flexibility; \textit{i.e.} it took $a_3 = a_4 = 0$. In addition, it used $n_2= \ddot r/r \Omega^2$
and  introduced a (real) modulus NQC factor $f_{\ell m}^{\rm NQC}$ only for the dominant quadrupolar wave $\ell = 2 = m$.] We henceforth use $M$ as time unit. This result relies on the proper comparison between NR and EOB time series, which is a delicate subject. In fact, to compare the NR and EOB phase time-series 
$ \phi_{22}^{\rm NR}(t_{\rm NR})$ and $\phi_{22}^{\rm EOB}(t_{\rm EOB})$
one needs to shift, by additive constants, both one of the time variables, and 
one of the phases. In other words, we need to determine $\tau$ and $\alpha$ such that
the ``shifted'' EOB quantities 
\begin{equation}
t'_{\rm EOB}=t_{\rm EOB} + \tau \ , \quad
\phi_{22}^{'\rm EOB} = \phi_{22}^{\rm EOB} + \alpha
\end{equation}
``best fit'' the NR ones. One convenient way to do so is first to ``pinch'' (\textit{i.e.} constrain to vanish) the EOB/NR phase difference at two different instants (corresponding to two
different frequencies $\omega_1$ and $\omega_2$). Having so related the EOB time and phase variables to the NR ones we can 
straigthforwardly compare the EOB time series to its NR correspondant.
In particular, we can compute the (shifted) EOB--NR phase difference
\begin{figure}[t]
\begin{center}
\hglue-12mm\includegraphics[height=5cm]{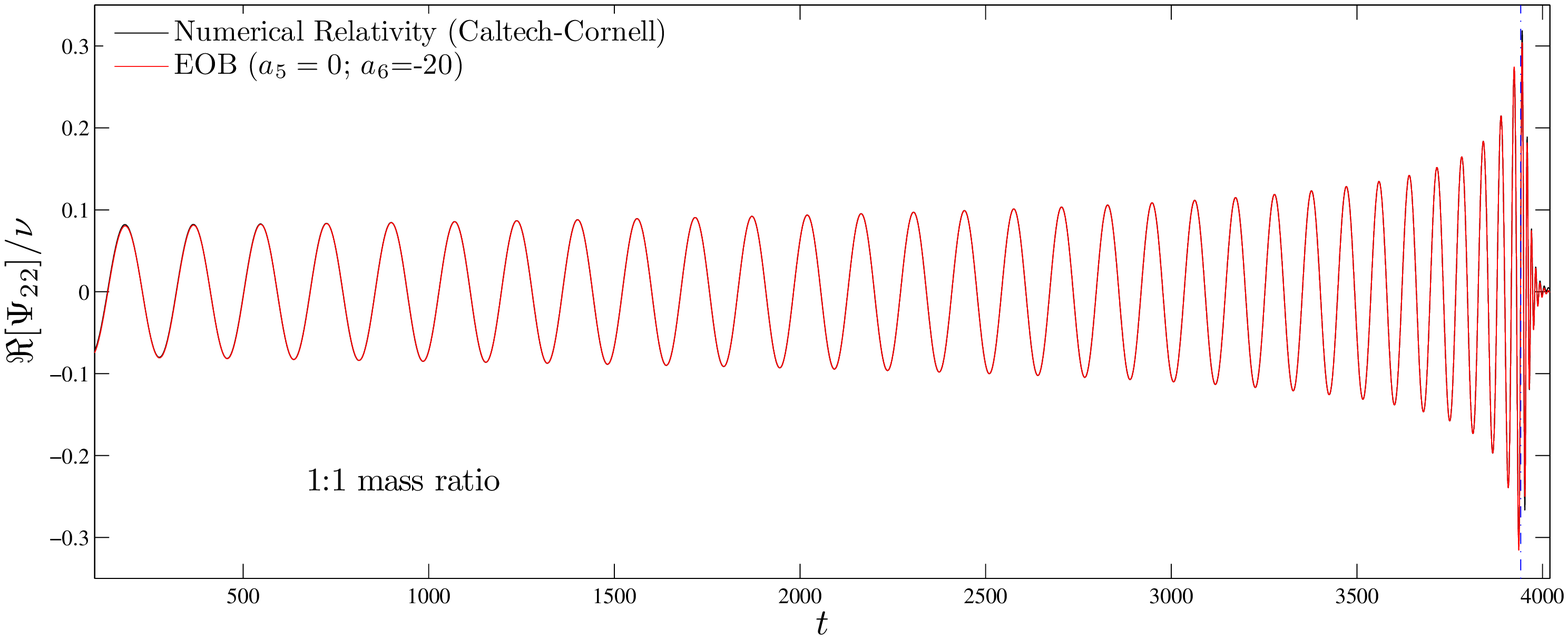}
\caption{This figure illustrates the comparison (made in Refs.~\cite{DamourNagar2009,DamourNagar2011}) between the (NR-completed) EOB waveform (Zerilli-normalized quadrupolar ($\ell=m=2$) metric waveform~(\ref{eqn18}) with parameter-free radiation reaction~(\ref{eq:RR_new}) and with $a_5=0$, $a_6=-20$) and one of the most accurate numerical relativity waveform (equal-mass case) nowadays available \cite{Scheel2009}. The phase difference between the two is $\Delta\phi\leq\pm 0.01$ radians during the entire inspiral and plunge, which is at the level of the numerical error.
\label{fig:waveform}}
\end{center}
\end{figure}

\begin{figure}[t]
\begin{center}
\includegraphics[height=7cm ]{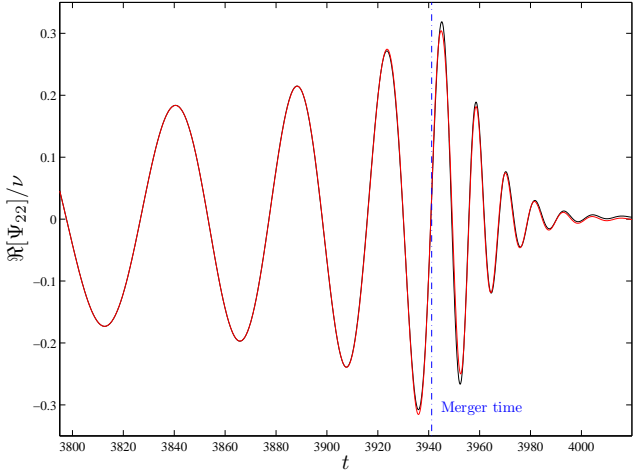}
\caption{\label{fig:ringdown}Close up around merger 
of the waveforms of Fig.~\ref{fig:waveform}. Note the excellent
agreement between \textit{both} modulus and phasing also during
the ringdown phase.}
\end{center}
\end{figure}
\begin{equation}
\label{deltaphi}
\Delta^{\omega_1,\omega_2}\phi_{22}^{\rm EOB NR} (t_{\rm NR}) \equiv \phi_{22}^{'\rm EOB}(t'^{\rm EOB}) - \phi_{22}^{\rm NR}(t^{\rm NR}).
\end{equation}
Figure~\ref{fig:waveform} compares\footnote{The two ``pinching'' frequencies used for
this comparison are $M\omega_1=0.047$ and $M\omega_2=0.31$.} 
(the real part of) the analytical EOB \textit{metric} quadrupolar waveform 
$\Psi^{\rm EOB}_{22}/\nu$ to the corresponding 
(Caltech-Cornell-CITA) NR \textit{metric} waveform 
$\Psi^{\rm NR}_{22}/\nu$. [Here, $\Psi_{22}$ denotes the Zerilli-normalized asymptotic quadrupolar waveform, \textit{i.e.} $\Psi_{22} \equiv \widehat R h_{22} / \sqrt{24}$ with $\widehat R = Rc^2 / GM$.] This NR  metric waveform has
been obtained by a double time-integration 
(following the procedure of Ref.~\cite{DNHHB2008})
from the original, publicly available, \textit{curvature}
waveform $\psi_4^{22}$ \cite{Scheel2009}. Such a curvature waveform has
been extrapolated \textit{both} in resolution and in
extraction radius. The agreement between the analytical prediction 
and the NR result is striking, even around the merger.
See Fig.~\ref{fig:ringdown} which closes up on the
merger. The vertical line indicates the
location of the EOB-merger time, i.e., the location
of the maximum of the orbital frequency.

\medskip

The phasing agreement between the waveforms is excellent 
over the full time span of the simulation
(which covers 32 cycles of inspiral and about 6
cycles of ringdown), while the modulus agreement is excellent
over the full span, apart from two cycles after merger
where one can notice a difference. 
More precisely, the phase difference, 
$\Delta \phi= \phi_{\rm metric}^{\rm EOB}-\phi_{\rm metric}^{\rm NR}$, 
remains remarkably small ($\sim \pm 0.02$ radians) during the
entire inspiral and plunge ($\omega_2=0.31$ being quite near
the merger). By comparison, the root-sum of 
the various numerical errors on the phase 
(numerical truncation, outer boundary, 
extrapolation to infinity) is about $0.023$ 
radians during the inspiral~\cite{Scheel2009}. 
At the merger,
and during the ringdown, $\Delta \phi$ takes somewhat 
larger values ($\sim \pm 0.1$ radians), but it oscillates around
zero, so that, on average, it stays very well in phase
with the NR waveform whose error rises to $\pm 0.05$ radians during ringdown. 
In addition, Ref.~\cite{DamourNagar2009} compared the 
EOB waveform to accurate numerical relativity data (obtained by
the Jena group~\cite{DNHHB2008}) on the coalescence of
\textit{unequal mass-ratio} black-hole binaries. Again, the agreement 
was good, and within the numerical error bars. 

\medskip

This type of high-accuracy comparison between NR waveforms and EOB[NR] ones (where EOB[NR] denotes a EOB formalism which has been completed by fitting some EOB-flexibility parameters to NR data) has been pursued and extended in Ref.~\cite{Pan2011}. The latter reference used the ``improved'' EOB formalism of Ref.~\cite{DamourNagar2009} with some variations (\textit{e.g.} a third modulus NQC coefficient $a_i$, two phase NQC coefficients, the argument $x_{\Omega} = (M \Omega)^{2/3}$ in $(\rho_{\ell m}^{\rm Taylor} (x))^{\ell}$, eight QNM modes) and calibrated it to NR simulations of mass ratios $q = m_2 / m_1 = 1,2,3,4$ and $6$ performed by the Caltech-Cornell-CITA
group~\cite{Buchman:2012dw,MacDonald:2012mp}. They considered not only the leading $(\ell , m) = (2,2)$ GW mode, but the subleading ones $(2,1), (3,3), (4,4)$ and $(5,5)$. They found that, for this large range of mass ratios, EOB[NR] (with suitably fitted, $\nu$-dependent values of $a_5$ and $a_6$) was able to describe the NR waveforms essentially within the NR errors. This confirms the usefulness of the EOB formalism in helping the detection and analysis of upcoming GW signals.

\medskip

Here, having in view GW observations from ground-based interferometric detectors we focussed on comparable-mass systems. The EOB formalism has also been compared to NR results in the extreme mass-ratio limit $\nu \ll 1$. In particular, Ref.~\cite{Bernuzzi2011} found an excellent agreement between the analytical and numerical results.

\subsection{EOB[3PN] dynamics vs NR one}

Let us also mention other types of EOB/NR comparisons. Recently, two examples of EOB/NR comparisons have been performed directly at the level of the \textit{dynamics} of a binary black hole, rather than at the level of the waveform. Moreover, contrary to the waveform comparisons of the previous subsection which involved an NR-completed EOB formalism (``EOB[NR]''), the dynamical comparisons we are going to discuss involve the purely analytical 3PN-accurate EOB formalism (``EOB[3PN]''), without any NR-based improvement.

\medskip

First, Le Tiec et al. \cite{LeTiec2011} have extracted from accurate NR simulations of slightly eccentric binary black-hole systems (for several mass ratios $q = m_1/m_2$ between $1/8$ and $1$) the function relating the periastron-advance parameter
$$
K = 1+\frac{\Delta\Phi}{2\pi} \, ,
$$ 
(where $\Delta\Phi$ is the periastron advance per radial period) to the dimensionless averaged angular frequency $M\Omega_{\varphi}$ (with $M = m_1 + m_2$ as above). Then they compared the NR-estimate of the mass-ratio dependent functional relation
$$
K = K (M\Omega_{\varphi} ; \nu) \, ,
$$
where $\nu = q/(1+q)^2$, to the predictions of various analytic approximation schemes: PN theory, EOB theory and two different ways of using GSF theory. Let us only mention here that the prediction from the purely analytical EOB[3PN] formalism for $K(M\Omega_{\varphi} ; \nu)$ \cite{Damour2010} agreed remarkably well (essentially within numerical errors) with its NR estimate for all mass ratios, while, by contrast, the PN-expanded prediction for $K(M\Omega_{\varphi} ; \nu)$ \cite{Damour2000} showed a much poorer agreement, especially as $q$ moved away from 1.

\medskip

Second, Damour, Nagar, Pollney and Reisswig \cite{Damour2012} have recently extracted from accurate NR simulations of black-hole binaries (with mass ratios $q = m_2 / m_1 = 1,2$ and $3$) the gauge-invariant relation between the (reduced) binding energy $E = ({\mathcal E}^{\rm tot} - M)/\mu$ and the (reduced) angular momentum $j = J / (G\mu M)$ of the system. Then they compared the NR-estimate of the mass-ratio dependent functional relation
$$
E = E(j;\nu)
$$
to the predictions of various analytic approximation schemes: PN theory and various versions of EOB theory (some of these versions were NR-completed). Let us only mention here that the prediction from the purely analytical, 3PN-accurate EOB[3PN] for $E(j;\nu)$ agreed remarkably well with its NR estimate (for all mass ratios) essentially \textit{down to the merger}. This is illustrated in Fig.~4 for the $q=1$ case. By contrast, the 3PN expansion in (powers of $1/c^2$) of the function $E(j;\nu)$ showed a much poorer agreement (for all mass ratios).

\begin{figure}[t]
\begin{center}
\includegraphics[height=7cm ]{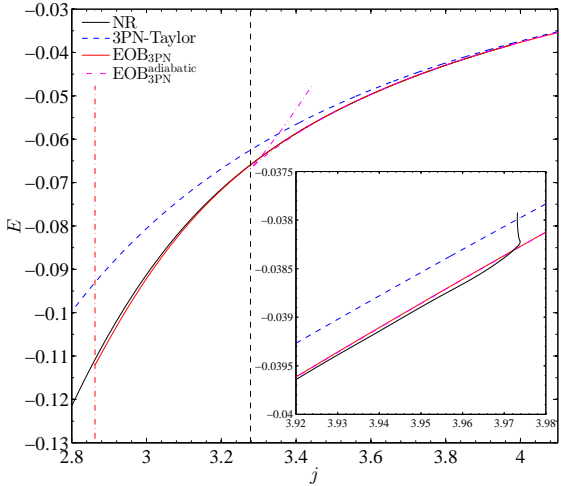}
\caption{\label{fig:??} Comparison (made in \cite{Damour2012}) between various analytical estimates of the energy-angular momentum functional relation and its numerical-relativity estimate (equal-mass case). The standard ``Taylor-expanded'' 3PN $E(j)$ curve shows the largest deviation from NR results, especially at low $j$'s, while the two (adiabatic and nonadiabatic) 3PN-accurate, \textit{non-NR-calibrated} EOB $E(j)$ curves agree remarkably well with the NR one.}
\end{center}
\end{figure}

\section{Other developments}

\subsection{EOB with spinning bodies}

We lack space here for discussing the extension of the EOB formalism to binary systems made of spinning bodies. Let us start by mentionning that the spin-extension of the EOB formalism was initiated in Ref.~\cite{TDamour2001}, that the first EOB-based analytical calculation of a complete waveform from a spinning binary was performed in Ref.~\cite{Buonanno2006}, and that the first attempt at calibrating a spinning EOB model to accurate NR simulations of spinning (non precessing) black-hole binaries was presented in \cite{Pan2010}. In addition, several formal aspects related to the inclusion of spins in the EOB formalism have been discussed in Refs.~\cite{DJS2008,Barausse2009,Barausse2010,Nagar2011,Barausse2011} (see references within these papers for PN works dealing with spin effects) and a generalization of the factorized multipolar waveform of Ref.~\cite{Damour2009} to spinning, non-precessing binaries has been constructed in Refs.~\cite{YPan2011,Taracchini2012}.

\subsection{EOB with tidally deformed bodies}

In binary systems comprising \textit{neutron stars}, rather than black holes, the tidal deformation of the neutron star(s) will significantly modify the phasing of the emitted gravitational waveform during the late inspiral. As GW's from binary neutron stars are expected sources for upcoming ground-based GW detectors, it is important to extend the EOB formalism by including tidal effects (see \cite{TDamour2012} and references therein). This extension has been defined in Refs.~\cite{TDamour2010,BDF2012}. The comparison between this tidal-extended EOB and state-of-the-art NR simulations of neutron-star binaries has been discussed in Refs.~\cite{Baiotti2010,Baiotti2011}. It appears from these comparisons that the tidal-extended EOB formalism is able to describe the motion and radiation of neutron-star binaries within NR errors. More accurate simulations will be needed to ascertain whether one needs to calibrate some higher-order flexibility parameters of the tidal-EOB formalism, or whether the currently known analytic accuracy is sufficient.

\subsection{EOB and GSF}

We mentioned in the Introduction that GSF theory has recently opened a new source of information on the general relativistic two-body problem. Let us briefly mention here that there has been, recently, a quite useful transfer of information from GSF theory to EOB theory. The program of using GSF-theory to improve EOB-theory was first highlighted in Ref.~\cite{Damour2010}. That work pointed to several concrete gauge-invariant calculations (within GSF theory) that would provide accurate information about the $O(\nu)$ contributions to several EOB potentials. More precisely, let us define the functions $a(u)$ and $\bar d (u)$ as the $\nu$-linear contributions to the EOB potentials $A(u;\nu)$ and $\overline D (u;\nu) \equiv D^{-1} (u;\nu)$:
$$
A(u;\nu) = 1-2u + \nu \, a(u) + O(\nu^2) \, ,
$$
$$
\overline D (u;\nu) = (AB)^{-1} = 1 + \nu \, \bar d(u) + O(\nu^2) \, .
$$
Ref.~\cite{Damour2010} has shown that a computation of the GSF-induced correction to the periastron advance of slightly eccentric orbits would allow one to compute the following combination of EOB functions
$$
\bar\rho (u) = a(u) + u \, a' (u) + \frac12 \, u (1-2u) \, a''(u) + (1-6u) \, \bar d (u) \, .
$$
The GSF-calculation of the EOB function $\bar\rho (u)$ was then performed in Ref.~\cite{Barack2010} (in the range $0 \leq u \leq \frac16$).

\medskip

More recently, a series of works by Le Tiec and collaborators \cite{LeTiec2012,ALeTiec2012,Barausse2012} have (through an indirect route) shown how GSF calculations could be used to compute the EOB $\nu$-linear $a(u)$ function separately from the $\bar d (u)$ one. Ref.~\cite{Barausse2012} then gave a fitting formula for $a(u)$ over the interval $0 \leq u \leq \frac15$ as well as accurate estimates of the coefficients of the Taylor expansion of $a(u)$ around $u=0$ (corresponding to the knowledge of the PN expansion of $a(u)$ to a very high PN order). Very recently, Ackay et al. \cite{Akcay2012} succeeded in accurately computing (through GSF theory) the EOB $a(u)$ function over the larger interval $0 \leq u \leq \frac13$. It was (surprisingly) found that $a(u)$ \textit{diverges} like $a (u) \approx 0.25 (1-3u)^{-1/2}$ at the light-ring limit $u \to \left( \frac13 \right)^-$. The meaning for EOB theory of this singular behavior of $a(u)$ at the light-ring is discussed in detail in Ref.~\cite{Akcay2012}.

\subsection{Toward further improvements to EOB}

Let us finally mention some avenues for further progress in EOB theory.

\medskip

Logarithmic contributions to the $A(u)$ and $\overline D (u)$ functions have been recently computed at the 4PN level \cite{Damour2010,Blanchet2010} and even the 5PN one \cite{Damourlogs,Barausse2012}. They have been incorporated in a recent, improved implementation of the EOB formalism \cite{DNB2012}.

\medskip

Two groups have embarked on a calculation of the (full) conservative dynamics at the 4PN level \cite{Foffa2012,Jaranowski2012}. If they succeed, it will be important to translate their \textit{gauge-dependent} results in the \textit{gauge-invariant} form used in EOB theory. [Remember that EOB theory is essentially based on the gauge-invariant Delaunay Hamiltonian $H(I_a)$.]

\medskip

More generally, let us emphasize that the EOB formalism provides a convenient, gauge-invariant way of packaging both the conservative dynamics and the multipolar waveform. This EOB packaging  has often turned out
to be very {\it economical}. We recommend that authors computing high-order PN corrections to either the dynamics or the waveform reexpress their results in terms of the EOB building blocks. 

For instance, 
Jaranowski and Scha\"efer \cite{Jaranowski2012} have recently given a partial result at 4PN, expressed in terms of
the (gauge-invariant) function $E(M\Omega_{\varphi} ; \nu)$. In terms of this function,
the 4PN contribution is a polynomial of the {\it fourth degree} in $\nu$, namely, with
$x \equiv (M\Omega_{\varphi} )^{2/3}$ and 
\begin{eqnarray*}
E(x;\nu) &=& -\frac12 \mu c^2 x  ( 1+e_{1PN}(\nu) x + e_{2PN}(\nu) x^2+ e_{3PN}(\nu) x^3  \\
&&  + e_{4PN}(\nu ; \ln x) x^4 + O(x^5 \ln x)  ),
\end{eqnarray*}
they found
\begin{equation} \label{e4PN}
e_{4PN}(\nu ; \ln x) = - \frac{3969}{128} + c_1 \nu + c_2 \nu^2 + \frac{301}{1728} \nu^3 + \frac{77}{31104} \nu^4 + \frac{448}{15} \nu \ln x,
\end{equation}
where they could not compute the values of the coefficients $c_1$ and $c_2$ of the
terms linear and quadratic in $\nu$, but only the contributions cubic and quartic in $\nu$.
We wish to point out that their result is re-expressed in a more economical (and more informative) way
in terms of the basic EOB potential $A(u;\nu)$. Indeed, in terms of the PN expansion,
 of $A(u;\nu)$,
$$
A^{\rm Taylor} (u;\nu) = 1-2u + \widetilde a_3 (\nu) u^3 + \widetilde a_4 (\nu) u^4 + \widetilde a_5 (\nu; \ln u) u^5 + \widetilde a_6 (\nu; \ln u) u^6 + \ldots
$$
the information contained in the above result  can be entirely re-expressed in terms of the 4PN-level
coefficient $\widetilde a_5 (\nu; \ln u)$. When doing this re-expression, one then finds that the information content
of  Eq.~(\ref{e4PN})  is  that the 4PN-level EOB coefficient $\widetilde a_5 (\nu; \ln u)$ is {\it no more than quadratic} in $\nu$, i.e.
$$
\widetilde a_5 (\nu; \ln u)= (a_5  + \frac{64}{5} \ln u)\nu + a'_5 \nu^2  \, ,
$$
{\it without} contributions of degree $\nu^3$ and $\nu^4$. We recall that similar
cancellations of  higher $\nu^n$ terms were found at lower PN orders in the EOB $A(u;\nu)$
function. Namely, they were found to contain only terms {\it linear} in $\nu$, while 
$\widetilde a_3 (\nu)$ could a priori have been quadratic in $\nu$, and $\widetilde a_4 (\nu)$ 
could a priori have been cubic in $\nu$.  The fact that similar remarkable cancellations
still hold, according to the result of \cite{Jaranowski2012},  at the 4PN level, is a clear
indication that the EOB packaging of information of the dynamics in the $A(u;nu)$
potential is quite compact. Indeed, it says that the two complicated 
terms $\frac{301}{1728} \nu^3 + \frac{77}{31104} \nu^4 $ in the energy function
are already encoded in the structure of the EOB formalism.
Finally, note that the full  gauge-invariant content  of a 4PN computation of the dynamics,
when interpreted within the EOB
formalism, is  described by only three EOB terms: the coefficient $\widetilde a_5 (\nu;\ln u)$ in $A(u;\nu)$, 
an analogous coefficient  $\widetilde{\!\!\bar d}_4 (\nu;\ln u)$ in $\bar D(u;\nu)$,
and an additional contribution to $Q(p)$.  

Regarding the waveform, let us mention  another recent example where it would have been
useful and clarifying to use the EOB packaging.
Namely, when re-expressing it in terms of
the factorized EOB waveform, the new content of the recent $3.5$PN level
computation by Faye,  Marsat,  Blanchet, and  Iyer \cite{Faye2012} of the PN-expanded quadrupolar waveform $h_{22}$,
 is entirely contained in an additional $3.5$PN-level contribution to the supplementary phase, 
 namely $\delta_{22} = (30995/1134\,\nu + 962/135\, \nu^2) \, x^{7/2}$. Indeed, the $3.5$PN-level contributions to the modulus computed in\cite{Faye2012}  were already included in the factorized EOB waveform of Ref.~\cite{DamourNagar2009}.

\section{Conclusions}

We hope that this brief review has made it clear that:

\medskip

1. There is a \textit{complementarity} between the various current approaches to the general relativistic two-body problem: post-Newtonian, Effective One Body, gravitational self-force and numerical relativity simulations (of both comparable-mass
and extreme-mass-ratio systems).

\medskip

2. The effective one body formalism offers a convenient framework for combining, in a synergetic manner, information coming from the other approaches. This formalism seems to constitute an efficient way to analytically describe the motion and radiation of circularized\footnote{See \cite{Bini2012} for a recent extension of the EOB formalism to non-circular 
(ellipticlike or hyperboliclike) motions.} binaries, and to provide accurate gravitational wave templates for detection and data analysis.

\medskip

3. The general relativistic two-body problem is more lively than ever. It provides an example of Poincar\'e's sentence: ``Il n'y a pas de probl\`emes r\'esolus, il y a seulement des probl\`emes plus ou moins r\'esolus.'' [``There are no (definitely) solved problems, there are only more-or-less solved problems.'']

\section*{References}
\bibliography{TheGRTBP}

\begin{thebibliography}{100}

\bibitem{Akcay2012}
Akcay, Sarp, Barack, Leor, Damour, Thibault  and Sago, Norichika,
  ``{Gravitational self-force and the effective-one-body formalism between the
  innermost stable circular orbit and the light ring}'', {\em Phys.Rev.}, {\bf
  D86}, 104041, (2012).
  {\small[\href{http://dx.doi.org/10.1103/PhysRevD.86.104041}{DOI}]},
  {\small[\href{http://arxiv.org/abs/1209.0964}{{arXiv:1209.0964
  {\small[gr-qc]}}}]}.

\bibitem{Baiotti2010}
Baiotti, L., Damour, T., Giacomazzo, B., Nagar, A.  and Rezzolla, L.,
  ``Analytic modelling of tidal effects in the relativistic inspiral of binary
  neutron stars'', {\em Phys. Rev. Lett.}, {\bf 105}, 261101, (2010).
  {\small[\href{http://arxiv.org/abs/1009.0521 [gr-qc]}{{1009.0521 [gr-qc]}}]}.

\bibitem{Baiotti2011}
Baiotti, L., Damour, T., Giacomazzo, B., Nagar, A.  and Rezzolla, L.,
  ``Accurate numerical simulations of inspiralling binary neutron stars and
  their comparison with effective-one-body analytical models'', {\em Phys. Rev.
  D}, {\bf 84}, 024017, (2011). {\small[\href{http://arxiv.org/abs/1103.3874
  [gr-qc]}{{1103.3874 [gr-qc]}}]}.

\bibitem{Baker2006}
Baker, J.G., Centrella, J., Choi, D.I., Koppitz, M.  and van Meter, J.,
  ``Gravitational wave extraction from an inspiraling configuration of merging
  black holes'', {\em Phys. Rev. Lett.}, {\bf 96}, 111102, (2006).
  {\small[\href{http://arxiv.org/abs/gr-qc/0511103}{{gr-qc/0511103}}]}.

\bibitem{Barack2009}
Barack, L., ``Gravitational self force in extreme mass-ratio inspirals'', {\em
  Class. Quant. Grav.}, {\bf 26}, 213001, (2009).
  {\small[\href{http://arxiv.org/abs/0908.1664 [gr-qc]}{{0908.1664 [gr-qc]}}]}.

\bibitem{Barack2010}
Barack, L., Damour, T.  and Sago, N., ``Precession effect of the gravitational
  self-force in a Schwarzschild spacetime and the effective one-body
  formalism'', {\em Phys. Rev. D}, {\bf 82}, 084036, (2010).
  {\small[\href{http://arxiv.org/abs/1008.0935 [gr-qc]}{{1008.0935 [gr-qc]}}]}.

\bibitem{Barausse2010}
Barausse, E.  and Buonanno, A., ``An Improved effective-one-body Hamiltonian
  for spinning black-hole binaries'', {\em Phys. Rev. D}, {\bf 81}, 084024,
  (2010). {\small[\href{http://arxiv.org/abs/0912.3517 [gr-qc]}{{0912.3517
  [gr-qc]}}]}.

\bibitem{Barausse2011}
Barausse, E.  and Buonanno, A., ``Extending the effective-one-body Hamiltonian
  of black-hole binaries to include next-to-next-to-leading spin-orbit
  couplings'', {\em Phys. Rev. D}, {\bf 84}, 104027, (2011).
  {\small[\href{http://arxiv.org/abs/1107.2904 [gr-qc]}{{1107.2904 [gr-qc]}}]}.

\bibitem{Barausse:2011kb}
Barausse, Enrico, Buonanno, Alessandra, Hughes, Scott~A., Khanna, Gaurav,
  O'Sullivan, Stephen  {et~al.}, ``{Modeling multipolar gravitational-wave
  emission from small mass-ratio mergers}'', {\em Phys.Rev.}, {\bf D85},
  024046, (2012).
  {\small[\href{http://dx.doi.org/10.1103/PhysRevD.85.024046}{DOI}]},
  {\small[\href{http://arxiv.org/abs/1110.3081}{{arXiv:1110.3081
  {\small[gr-qc]}}}]}.

\bibitem{Barausse2012}
Barausse, E., Buonanno, A.  and Le~Tiec, A., ``The complete non-spinning
  effective-one-body metric at linear order in the mass ratio'', {\em Phys.
  Rev. D}, {\bf 85}, 064010, (2012).
  {\small[\href{http://arxiv.org/abs/1111.5610 [gr-qc]}{{1111.5610 [gr-qc]}}]}.

\bibitem{Barausse2009}
Barausse, E., Racine, E.  and Buonanno, A., ``Hamiltonian of a spinning
  test-particle in curved spacetime. [Erratum-ibid. D {\bf 85}, 069904
  (2012)]'', {\em Phys. Rev. D}, {\bf 80}, 104025, (2009).
  {\small[\href{http://arxiv.org/abs/0907.4745 [gr-qc]}{{0907.4745 [gr-qc]}}]}.

\bibitem{Bernuzzi2011}
Bernuzzi, S., Nagar, A.  and Zenginoglu, A., ``Binary black hole coalescence in
  the extreme-mass-ratio limit: testing and improving the effective-one-body
  multipolar waveform'', {\em Phys. Rev. D}, {\bf 83}, 064010, (2011).
  {\small[\href{http://arxiv.org/abs/1012.2456 [gr-qc]}{{1012.2456 [gr-qc]}}]}.

\bibitem{SBernuzzi2011}
Bernuzzi, S., Nagar, A.  and Zenginoglu, A., ``Binary black hole coalescence in
  the large-mass-ratio limit: the hyperboloidal layer method and waveforms at
  null infinity'', {\em Phys. Rev. D}, {\bf 84}, 084026, (2011).
  {\small[\href{http://arxiv.org/abs/1107.5402 [gr-qc]}{{1107.5402 [gr-qc]}}]}.

\bibitem{Berti2007}
Berti, E., Cardoso, V., Gonzalez, J.A., Sperhak, U., Hannam, M., Husa, S.  and
  Bruegmann, B., ``Inspiral, merger and ringdown of unequal mass black hole
  binaries: A multipolar analysis'', {\em Phys. Rev. D}, {\bf 76}, 064034,
  (2007). {\small[\href{http://arxiv.org/abs/gr-qc/0703053}{{gr-qc/0703053}}]}.

\bibitem{Bini2012}
Bini, D.  and Damour, T., ``Gravitational radiation reaction along general
  orbits in the effective one-body formalism'', {\em Phys. Rev. D}, {\bf 86},
  124012, (2012).
  {\small[\href{http://arxiv.org/abs/1210.2834[gr-qc]}{{1210.2834[gr-qc]}}]}.

\bibitem{BDF2012}
Bini, D., Damour, T.  and Faye, G., ``Effective action approach to higher-order
  relativistic tidal interactions in binary systems and their effective one
  body description'', {\em Phys. Rev. D}, {\bf 85}, 124034, (2012).
  {\small[\href{http://arxiv.org/abs/1202.3565 [gr-qc]}{{1202.3565 [gr-qc]}}]}.

\bibitem{Blanchet1995}
Blanchet, L., ``Second Postnewtonian Generation Of Gravitational Radiation'',
  {\em Phys. Rev. D}, {\bf 51}, 2559, (1995).
  {\small[\href{http://arxiv.org/abs/gr-qc/9501030}{{gr-qc/9501030}}]}.

\bibitem{Blanchet1998}
Blanchet, L., ``Gravitational-wave tails of tails, [Erratum-ibid.\ {\bf 22},
  3381 (2005)]'', {\em Class. Quant. Grav.}, {\bf 15}, 113, (1998).
  {\small[\href{http://arxiv.org/abs/gr-qc/9710038}{{gr-qc/9710038}}]}.

\bibitem{LBlanchet1998}
Blanchet, L., ``Quadrupole-quadrupole gravitational waves'', {\em Class. Quant.
  Grav.}, {\bf 15}, 89, (1998).
  {\small[\href{http://arxiv.org/abs/gr-qc/9710037}{{gr-qc/9710037}}]}.

\bibitem{LBlanchet2002}
Blanchet, L., ``Gravitational radiation from post-Newtonian sources and
  inspiralling compact binaries'', {\em Living Rev. Rel.}, {\bf 5}, 3, (2002).
  {\small[\href{http://arxiv.org/abs/gr-qc/0202016}{{gr-qc/0202016}}]}.

\bibitem{Blanchet1986}
Blanchet, L.  and Damour, T., ``Radiative gravitational fields in general
  relativity I. General structure of the field outside the source'', {\em Phil.
  Trans. Roy. Soc. Lond. A}, {\bf 320}, 379, (1986).

\bibitem{Blanchet1989}
Blanchet, L.  and Damour, T., ``Postnewtonian generation of gravitational
  waves'', {\em Annales Poincar\'e Phys. Theor.}, {\bf 50}, 377, (1989).

\bibitem{Blanchet1992}
Blanchet, L.  and Damour, T., ``Hereditary Effects In Gravitational
  Radiation'', {\em Phys. Rev. D}, {\bf 46}, 4304, (1992).

\bibitem{Blanchet2004}
Blanchet, L., Damour, T.  and Esposito-Far\`ese, G., ``Dimensional
  regularization of the third post-Newtonian dynamics of point particles in
  harmonic coordinates'', {\em Phys. Rev. D}, {\bf 69}, 124007, (2004).
  {\small[\href{http://arxiv.org/abs/gr-qc/0311052}{{gr-qc/0311052}}]}.

\bibitem{BDEI2004}
Blanchet, L., Damour, T., Esposito-Far\`ese, G.  and Iyer, B.R.,
  ``Gravitational radiation from inspiralling compact binaries completed at the
  third post-Newtonian order'', {\em Phys. Rev. Lett.}, {\bf 93}, 091101,
  (2004). {\small[\href{http://arxiv.org/abs/gr-qc/0406012}{{gr-qc/0406012}}]}.

\bibitem{BDEI2005}
Blanchet, L., Damour, T., Esposito-Far\`ese, G.  and Iyer, B.R., ``Dimensional
  regularization of the third post-Newtonian gravitational wave generation from
  two point masses'', {\em Phys. Rev. D}, {\bf 71}, 124004, (2005).
  {\small[\href{http://arxiv.org/abs/gr-qc/0503044}{{gr-qc/0503044}}]}.

\bibitem{BDI1995}
Blanchet, L., Damour, T.  and Iyer, B.R., ``Gravitational Radiation Damping Of
  Compact Binary Systems To Second Postnewtonian Order, [Erratum-ibid.\ D {\bf
  54}, 1860 (1996)]'', {\em Phys. Rev. D}, {\bf 51}, 5360, (1995).
  {\small[\href{http://arxiv.org/abs/gr-qc/9501029}{{gr-qc/9501029}}]}.

\bibitem{BDIWW1995}
Blanchet, L., Damour, T., Iyer, B.R., Will, C.M.  and Wiseman, A.G.,
  ``Gravitational Radiation Damping Of Compact Binary Systems To Second
  Postnewtonian Order'', {\em Phys. Rev. Lett.}, {\bf 74}, 3515, (1995).
  {\small[\href{http://arxiv.org/abs/gr-qc/9501027}{{gr-qc/9501027}}]}.

\bibitem{Blanchet2010}
Blanchet, L., Detweiler, S.L., Le~Tiec, A.  and Whiting, B.F., ``High-Order
  Post-Newtonian Fit of the Gravitational Self-Force for Circular Orbits in the
  Schwarzschild Geometry'', {\em Phys. Rev. D}, {\bf 81}, 084033, (2010).
  {\small[\href{http://arxiv.org/abs/1002.0726 [gr-qc]}{{1002.0726 [gr-qc]}}]}.

\bibitem{Blanchet2001}
Blanchet, L.  and Faye, G., ``General relativistic dynamics of compact binaries
  at the third post-Newtonian order'', {\em Phys. Rev. D}, {\bf 63}, 062005,
  (2001). {\small[\href{http://arxiv.org/abs/gr-qc/0007051}{{gr-qc/0007051}}]}.

\bibitem{BFIS2008}
Blanchet, L., Faye, G., Iyer, B.R.  and Sinha, S., ``The third post-Newtonian
  gravitational wave polarisations and associated spherical harmonic modes for
  inspiralling compact binaries in quasi-circular orbits'', {\em Class. Quant.
  Grav.}, {\bf 25}, 165003, (2008).
  {\small[\href{http://arxiv.org/abs/0802.1249 [gr-qc]}{{0802.1249 [gr-qc]}}]}.

\bibitem{Blanchet2005}
Blanchet, L.  and Iyer, B.R., ``Hadamard regularization of the third
  post-Newtonian gravitational wave generation of two point masses'', {\em
  Phys. Rev. D}, {\bf 71}, 024004, (2005).
  {\small[\href{http://arxiv.org/abs/gr-qc/0409094}{{gr-qc/0409094}}]}.

\bibitem{Blanchet2002}
Blanchet, L., Iyer, B.R.  and Joguet, B., ``Gravitational waves from
  inspiralling compact binaries: Energy flux to third post-Newtonian order,
  [Erratum-ibid.\ D {\bf 71}, 129903 (2005)]'', {\em Phys. Rev. D}, {\bf 65},
  064005, (2002).
  {\small[\href{http://arxiv.org/abs/gr-qc/0105098}{{gr-qc/0105098}}]}.

\bibitem{Blanchet1993}
Blanchet, L.  and Sch\"afer, G., ``Gravitational wave tails and binary star
  systems'', {\em Class. Quant. Grav.}, {\bf 10}, 2699, (1993).

\bibitem{Boyle2007}
Boyle, M.~{\it et al.}, ``High-accuracy comparison of numerical relativity
  simulations with post-Newtonian expansions'', {\em Phys. Rev. D}, {\bf 76},
  124038, (2007). {\small[\href{http://arxiv.org/abs/0710.0158
  [gr-qc]}{{0710.0158 [gr-qc]}}]}.

\bibitem{Brezin1970}
Br\'ezin, E., Itzykson, C.  and Zinn-Justin, J., ``Relativistic balmer formula
  including recoil effects'', {\em Phys. Rev. D}, {\bf 1}, 2349, (1970).

\bibitem{Buchman:2012dw}
Buchman, Luisa~T., Pfeiffer, Harald~P., Scheel, Mark~A.  and Szilagyi, Bela,
  ``{Simulations of non-equal mass black hole binaries with spectral
  methods}'', {\em Phys.Rev.}, {\bf D86}, 084033, (2012).
  {\small[\href{http://dx.doi.org/10.1103/PhysRevD.86.084033}{DOI}]},
  {\small[\href{http://arxiv.org/abs/1206.3015}{{arXiv:1206.3015
  {\small[gr-qc]}}}]}.

\bibitem{Buonanno2006}
Buonanno, A., Chen, Y.  and Damour, T., ``Transition from inspiral to plunge in
  precessing binaries of spinning black holes'', {\em Phys. Rev. D}, {\bf 74},
  104005, (2006).
  {\small[\href{http://arxiv.org/abs/gr-qc/0508067}{{gr-qc/0508067}}]}.

\bibitem{Buonanno2007}
Buonanno, A., Cook, G.B.  and Pretorius, F., ``Inspiral, merger and ring-down
  of equal-mass black-hole binaries'', {\em Phys. Rev. D}, {\bf 75}, 124018,
  (2007). {\small[\href{http://arxiv.org/abs/gr-qc/0610122}{{gr-qc/0610122}}]}.

\bibitem{Buonanno1999}
Buonanno, A.  and Damour, T., ``Effective one-body approach to general
  relativistic two-body dynamics'', {\em Phys. Rev. D}, {\bf 59}, 084006,
  (1999). {\small[\href{http://arxiv.org/abs/gr-qc/9811091}{{gr-qc/9811091}}]}.

\bibitem{Buonanno2000}
Buonanno, A.  and Damour, T., ``Transition from inspiral to plunge in binary
  black hole coalescences'', {\em Phys. Rev. D}, {\bf 62}, 064015, (2000).
  {\small[\href{http://arxiv.org/abs/gr-qc/0001013}{{gr-qc/0001013}}]}.

\bibitem{ABuonanno2007}
Buonanno, A., Pan, Y., Baker, J.G., Centrella, J., Kelly, B.J., McWilliams,
  S.T.  and van Meter, J.R., ``Toward faithful templates for non-spinning
  binary black holes using the effective-one-body approach'', {\em Phys. Rev.
  D}, {\bf 76}, 104049, (2007). {\small[\href{http://arxiv.org/abs/0706.3732
  [gr-qc]}{{0706.3732 [gr-qc]}}]}.

\bibitem{Buonanno2009}
Buonanno, A., Pan, Y., Pfeiffer, H.P., Scheel, M.A., Buchman, L.T.  and Kidder,
  L.E., ``Effective-one-body waveforms calibrated to numerical relativity
  simulations: coalescence of non-spinning, equal-mass black holes'', {\em
  Phys. Rev. D}, {\bf 79}, 124028, (2009).
  {\small[\href{http://arxiv.org/abs/0902.0790 [gr-qc]}{{0902.0790 [gr-qc]}}]}.

\bibitem{Campanelli2006}
Campanelli, M., Lousto, C.O., Marronetti, P.  and Zlochower, Y., ``Accurate
  Evolutions of Orbiting Black-Hole Binaries Without Excision'', {\em Phys.
  Rev. Lett.}, {\bf 96}, 111101, (2006).
  {\small[\href{http://arxiv.org/abs/gr-qc/0511048}{{gr-qc/0511048}}]}.

\bibitem{DamourLH}
Damour, T., ``Gravitational radiation and the motion of compact bodies'', in
  Deruelle, N.  and Piran, T., eds., {\em Gravitational Radiation}, pp.
  59--144, (North-Holland, Amsterdam, 1983).

\bibitem{Damour1987}
Damour, T., ``The problem of motion in Newtonian and Einsteinian gravity'', in
  Hawking, S.W.  and Israel, W., eds., {\em Three Hundred Years of
  Gravitation}, pp. 128--198, (Cambridge University Press, Cambridge, 1987).

\bibitem{TDamour2001}
Damour, T., ``Coalescence of two spinning black holes: An effective one-body
  approach'', {\em Phys. Rev. D}, {\bf 64}, 124013, (2001).
  {\small[\href{http://arxiv.org/abs/gr-qc/0103018}{{gr-qc/0103018}}]}.

\bibitem{Damour2010}
Damour, T., ``Gravitational Self Force in a Schwarzschild Background and the
  Effective One Body Formalism'', {\em Phys. Rev. D}, {\bf 81}, 024017, (2010).
  {\small[\href{http://arxiv.org/abs/0910.5533 [gr-qc]}{{0910.5533 [gr-qc]}}]}.

\bibitem{Damourlogs}
Damour, T., ``(unpublished); cited in Ref.~\cite{Barack2010}, which quoted and
  used some combinations of the logarithmic contributions to $a(u)$ and $\bar d
  (u)$'', (2010).

\bibitem{Damour:1992we}
Damour, Thibault  and Esposito-Farese, Gilles, ``{Tensor multiscalar theories
  of gravitation}'', {\em Class.Quant.Grav.}, {\bf 9}, 2093--2176, (1992).
  {\small[\href{http://dx.doi.org/10.1088/0264-9381/9/9/015}{DOI}]}.

\bibitem{Damour:1995kt}
Damour, Thibault  and Esposito-Farese, Gilles, ``{Testing gravity to second
  postNewtonian order: A Field theory approach}'', {\em Phys.Rev.}, {\bf D53},
  5541--5578, (1996).
  {\small[\href{http://dx.doi.org/10.1103/PhysRevD.53.5541}{DOI}]},
  {\small[\href{http://arxiv.org/abs/gr-qc/9506063}{{arXiv:gr-qc/9506063
  {\small[gr-qc]}}}]}.

\bibitem{Damour:1998jk}
Damour, Thibault  and Esposito-Farese, Gilles, ``{Gravitational wave versus
  binary - pulsar tests of strong field gravity}'', {\em Phys.Rev.}, {\bf D58},
  042001, (1998).
  {\small[\href{http://dx.doi.org/10.1103/PhysRevD.58.042001}{DOI}]},
  {\small[\href{http://arxiv.org/abs/gr-qc/9803031}{{arXiv:gr-qc/9803031
  {\small[gr-qc]}}}]}.

\bibitem{Damour2006}
Damour, T.  and Gopakumar, A., ``Gravitational recoil during binary black hole
  coalescence using the effective one body approach'', {\em Phys. Rev. D}, {\bf
  73}, 124006, (2006).
  {\small[\href{http://arxiv.org/abs/gr-qc/0602117}{{gr-qc/0602117}}]}.

\bibitem{Damour2002}
Damour, T., Gourgoulhon, E.  and Grandcl\'ement, P., ``Circular orbits of
  corotating binary black holes: Comparison between analytical and numerical
  results'', {\em Phys. Rev. D}, {\bf 66}, 024007, (2002).
  {\small[\href{http://arxiv.org/abs/gr-qc/0204011}{{gr-qc/0204011}}]}.

\bibitem{Damour1991}
Damour, T.  and Iyer, B.R., ``Multipole analysis for electromagnetism and
  linearized gravity with irreducible cartesian tensors'', {\em Phys. Rev. D},
  {\bf 43}, 3259, (1991).

\bibitem{DamourIyer1991}
Damour, T.  and Iyer, B.R., ``PostNewtonian generation of gravitational waves.
  2. The Spin moments'', {\em Annales Poincar\'e Phys. Theor.}, {\bf 54}, 115,
  (1991).

\bibitem{Damour2009}
Damour, T., Iyer, B.R.  and Nagar, A., ``Improved resummation of post-Newtonian
  multipolar waveforms from circularized compact binaries'', {\em Phys. Rev.
  D}, {\bf 79}, 064004, (2009). {\small[\href{http://arxiv.org/abs/0811.2069
  [gr-qc]}{{0811.2069 [gr-qc]}}]}.

\bibitem{Damour1998}
Damour, T., Iyer, B.R.  and Sathyaprakash, B.S., ``Improved filters for
  gravitational waves from inspiralling compact binaries'', {\em Phys. Rev. D},
  {\bf 57}, 885, (1998).
  {\small[\href{http://arxiv.org/abs/gr-qc/9708034}{{gr-qc/9708034}}]}.

\bibitem{TDamour2000}
Damour, T., Jaranowski, P.  and Scha\"afer, G., ``Poincar\'e invariance in the
  ADM Hamiltonian approach to the general relativistic two-body problem,
  [Erratum-ibid. D {\bf 63}, 029903 (2001)]'', {\em Phys. Rev. D}, {\bf 62},
  021501, (2000).
  {\small[\href{http://arxiv.org/abs/gr-qc/0003051}{{gr-qc/0003051}}]}.

\bibitem{Damour2000}
Damour, T., Jaranowski, P.  and Sch\"afer, G., ``Dynamical invariants for
  general relativistic two-body systems at the third postNewtonian
  approximation'', {\em Phys. Rev. D}, {\bf 62}, 044024, (2000).
  {\small[\href{http://arxiv.org/abs/gr-qc/9912092}{{gr-qc/9912092}}]}.

\bibitem{DJS2000}
Damour, T., Jaranowski, P.  and Sch\"afer, G., ``On the determination of the
  last stable orbit for circular general relativistic binaries at the third
  post-Newtonian approximation'', {\em Phys. Rev. D}, {\bf 62}, 084011, (2000).
  {\small[\href{http://arxiv.org/abs/gr-qc/0005034}{{gr-qc/0005034}}]}.

\bibitem{Damour2001}
Damour, T., Jaranowski, P.  and Sch\"afer, G., ``Dimensional regularization of
  the gravitational interaction of point masses'', {\em Phys. Lett. B}, {\bf
  513}, 147--155, (2001).
  {\small[\href{http://arxiv.org/abs/gr-qc/0105038}{{gr-qc/0105038}}]}.

\bibitem{DJS2008}
Damour, T., Jaranowski, P.  and Sch\"afer, G., ``Effective one body approach to
  the dynamics of two spinning black holes with next-to-leading order
  spin-orbit coupling'', {\em Phys. Rev. D}, {\bf 78}, 024009, (2008).
  {\small[\href{http://arxiv.org/abs/0803.0915 [gr-qc]}{{0803.0915 [gr-qc]}}]}.

\bibitem{Damour2007}
Damour, T.  and Nagar, A., ``Faithful Effective-One-Body waveforms of
  small-mass-ratio coalescing black-hole binaries'', {\em Phys. Rev. D}, {\bf
  76}, 064028, (2007). {\small[\href{http://arxiv.org/abs/0705.2519
  [gr-qc]}{{0705.2519 [gr-qc]}}]}.

\bibitem{DamourNagar2007}
Damour, T.  and Nagar, A., ``Final spin of a coalescing black-hole binary: an
  Effective-One-Body approach'', {\em Phys. Rev. D}, {\bf 76}, 044003, (2007).
  {\small[\href{http://arxiv.org/abs/0704.3550 [gr-qc]}{{0704.3550 [gr-qc]}}]}.

\bibitem{Damour2008}
Damour, T.  and Nagar, A., ``Comparing Effective-One-Body gravitational
  waveforms to accurate numerical data'', {\em Phys. Rev. D}, {\bf 77}, 024043,
  (2008). {\small[\href{http://arxiv.org/abs/0711.2628 [gr-qc]}{{0711.2628
  [gr-qc]}}]}.

\bibitem{DamourNagar2009}
Damour, T.  and Nagar, A., ``An improved analytical description of inspiralling
  and coalescing black-hole binaries'', {\em Phys. Rev. D}, {\bf 79}, 081503,
  (2009). {\small[\href{http://arxiv.org/abs/0902.0136 [gr-qc]}{{0902.0136
  [gr-qc]}}]}.

\bibitem{TDamour2010}
Damour, T.  and Nagar, A., ``Effective One Body description of tidal effects in
  inspiralling compact binaries'', {\em Phys. Rev. D}, {\bf 81}, 084016,
  (2010). {\small[\href{http://arxiv.org/abs/0911.5041 [gr-qc]}{{0911.5041
  [gr-qc]}}]}.

\bibitem{DamourNagar2011}
Damour, T.  and Nagar, A., ``The Effective One Body description of the Two-Body
  problem'', {\em Fundam. Theor. Phys.}, {\bf 162}, 211, (2011).
  {\small[\href{http://arxiv.org/abs/0906.1769 [gr-qc]}{{0906.1769 [gr-qc]}}]}.

\bibitem{DNB2012}
Damour, T., Nagar, A.  and Bernuzzi, S., ``to be submitted for publication'',
  (2012).

\bibitem{DNHHB2008}
Damour, T., Nagar, A., Hannam, M., Husa, S.  and Bruegmann, B., ``Accurate
  Effective-One-Body waveforms of inspiralling and coalescing black-hole
  binaries'', {\em Phys. Rev. D}, {\bf 78}, 044039, (2008).
  {\small[\href{http://arxiv.org/abs/0803.3162 [gr-qc]}{{0803.3162 [gr-qc]}}]}.

\bibitem{DNNPR2008}
Damour, T., Nagar, A., Nils~Dorband, E., Pollney, D.  and Rezzolla, L.,
  ``Faithful Effective-One-Body waveforms of equal-mass coalescing black-hole
  binaries'', {\em Phys. Rev. D}, {\bf 77}, 084017, (2008).
  {\small[\href{http://arxiv.org/abs/0712.3003 [gr-qc]}{{0712.3003 [gr-qc]}}]}.

\bibitem{Damour2012}
Damour, T., Nagar, A., Pollney, D.  and Reisswig, C., ``Energy versus Angular
  Momentum in Black Hole Binaries'', {\em Phys. Rev. Lett.}, {\bf 108}, 131101,
  (2012). {\small[\href{http://arxiv.org/abs/1110.2938 [gr-qc]}{{1110.2938
  [gr-qc]}}]}.

\bibitem{Nagar2007}
Damour, T., Nagar, A.  and Tartaglia, A., ``Binary black hole merger in the
  extreme mass ratio limit'', {\em Class. Quant. Grav.}, {\bf 24}, S109,
  (2007). {\small[\href{http://arxiv.org/abs/gr-qc/0612096}{{gr-qc/0612096}}]}.

\bibitem{TDamour2012}
Damour, T., Nagar, A.  and Villain, L., ``Measurability of the tidal
  polarizability of neutron stars in late-inspiral gravitational-wave
  signals'', {\em Phys. Rev. D}, {\bf 85}, 123007, (2012).
  {\small[\href{http://arxiv.org/abs/1203.4352 [gr-qc]}{{1203.4352 [gr-qc]}}]}.

\bibitem{Damour1988}
Damour, T.  and Sch\"afer, G., ``Higher order relativistic periastron advances
  and binary pulsars'', {\em Nuovo Cim. B}, {\bf 101}, 127, (1988).

\bibitem{Davis1972}
Davis, M., Ruffini, R.  and Tiomno, J., ``Pulses of gravitational radiation of
  a particle falling radially into a Schwarzschild black hole'', {\em Phys.
  Rev. D}, {\bf 5}, 2932, (1972).

\bibitem{Epstein1975}
Epstein, R.  and Wagoner, R.V., ``Post-Newtonian generation of gravitational
  waves'', {\em Astrophys. J.}, {\bf 197}, 717--723, (1975).

\bibitem{Faye2012}
Faye, G., Marsat, S., Blanchet, L.  and Iyer, B.R., ``The third and a half
  post-Newtonian gravitational wave quadrupole mode for quasi-circular
  inspiralling compact binaries'', {\em Class. Quant. Grav.}, {\bf 29}, 175004,
  (2012). {\small[\href{http://arxiv.org/abs/1204.1043 [gr-qc]}{{1204.1043
  [gr-qc]}}]}.

\bibitem{Foffa2011}
Foffa, S.  and Sturani, R., ``Effective field theory calculation of
  conservative binary dynamics at third post-Newtonian order'', {\em Phys. Rev.
  D}, {\bf 84}, 044031, (2011). {\small[\href{http://arxiv.org/abs/1104.1122
  [gr-qc]}{{1104.1122 [gr-qc]}}]}.

\bibitem{Foffa2012}
Foffa, S.  and Sturani, R., ``The dynamics of the gravitational two-body
  problem in the post-Newtonian approximation at quadratic order in the
  Newton's constant'', (2012). {\small[\href{http://arxiv.org/abs/1206.7087
  [gr-qc]}{{1206.7087 [gr-qc]}}]}.

\bibitem{Fujita2012}
Fujita, R., ``Gravitational radiation for extreme mass ratio inspirals to the
  14th post-Newtonian order'', {\em Prog. Theor. Phys.}, {\bf 127}, 583,
  (2012). {\small[\href{http://arxiv.org/abs/1104.5615 [gr-qc]}{{1104.5615
  [gr-qc]}}]}.

\bibitem{Fujita2010}
Fujita, R.  and Iyer, B.R., ``Spherical harmonic modes of 5.5 post-Newtonian
  gravitational wave polarisations and associated factorised resummed waveforms
  for a particle in circular orbit around a Schwarzschild black hole'', {\em
  Phys. Rev. D}, {\bf 82}, 044051, (2010).
  {\small[\href{http://arxiv.org/abs/1005.2266 [gr-qc]}{{1005.2266 [gr-qc]}}]}.

\bibitem{Goldberger:2009qd}
Goldberger, Walter~D.  and Ross, Andreas, ``{Gravitational radiative
  corrections from effective field theory}'', {\em Phys.Rev.}, {\bf D81},
  124015, (2010).
  {\small[\href{http://dx.doi.org/10.1103/PhysRevD.81.124015}{DOI}]},
  {\small[\href{http://arxiv.org/abs/0912.4254}{{arXiv:0912.4254
  {\small[gr-qc]}}}]}.

\bibitem{Goldberger:2004jt}
Goldberger, Walter~D.  and Rothstein, Ira~Z., ``{An Effective field theory of
  gravity for extended objects}'', {\em Phys.Rev.}, {\bf D73}, 104029, (2006).
  {\small[\href{http://dx.doi.org/10.1103/PhysRevD.73.104029}{DOI}]},
  {\small[\href{http://arxiv.org/abs/hep-th/0409156}{{arXiv:hep-th/0409156
  {\small[hep-th]}}}]}.

\bibitem{Gonzalez2007}
Gonzalez, J.A., Sperhake, U., Bruegmann, B., Hannam, M.  and Husa, S., ``Total
  recoil: the maximum kick from nonspinning black-hole binary inspiral'', {\em
  Phys. Rev. Lett.}, {\bf 98}, 091101, (2007).
  {\small[\href{http://arxiv.org/abs/gr-qc/0610154}{{gr-qc/0610154}}]}.

\bibitem{MHannam2008}
Hannam, M., Husa, S., Bruegmann, B.  and Gopakumar, A., ``Comparison between
  numerical-relativity and post-Newtonian waveforms from spinning binaries: the
  orbital hang-up case'', {\em Phys. Rev. D}, {\bf 78}, 104007, (2008).
  {\small[\href{http://arxiv.org/abs/0712.3787 [gr-qc]}{{0712.3787 [gr-qc]}}]}.

\bibitem{Hannam2008}
Hannam, M., Husa, S., Sperhake, U., Bruegmann, B.  and Gonzalez, J.A., ``Where
  post-Newtonian and numerical-relativity waveforms meet'', {\em Phys. Rev. D},
  {\bf 77}, 044020, (2008). {\small[\href{http://arxiv.org/abs/0706.1305
  [gr-qc]}{{0706.1305 [gr-qc]}}]}.

\bibitem{Itoh2003}
Itoh, Y.  and Futamase, T., ``New derivation of a third post-Newtonian equation
  of motion for relativistic compact binaries without ambiguity'', {\em Phys.
  Rev. D}, {\bf 68}, 121501, (2003).
  {\small[\href{http://arxiv.org/abs/gr-qc/0310028}{{gr-qc/0310028}}]}.

\bibitem{Jaranowski1998}
Jaranowski, P.  and Sch\"afer, G., ``3rd post-Newtonian higher order Hamilton
  dynamics for two-body point-mass systems ([Erratum-ibid.\ D {\bf 63}, 029902
  (2001)])'', {\em Phys. Rev. D}, {\bf 57}, 72--74, (1998).
  {\small[\href{http://arxiv.org/abs/gr-qc/9712075}{{gr-qc/9712075}}]}.

\bibitem{Jaranowski2012}
Jaranowski, P.  and Sch\"afer, G., ``Towards the 4th post-Newtonian Hamiltonian
  for two-point-mass systems'', {\em Phys. Rev. D}, {\bf 86}, 061503, (2012).
  {\small[\href{http://arxiv.org/abs/1207.5448 [gr-qc]}{{1207.5448 [gr-qc]}}]}.

\bibitem{Kidder2008}
Kidder, L.E., ``Using Full Information When Computing Modes of Post-Newtonian
  Waveforms From Inspiralling Compact Binaries in Circular Orbit'', {\em Phys.
  Rev. D}, {\bf 77}, 044016, (2008).
  {\small[\href{http://arxiv.org/abs/0710.0614 [gr-qc]}{{0710.0614 [gr-qc]}}]}.

\bibitem{Konigsdorffer2003}
Konigsdorffer, C., Faye, G.  and Sch\"afer, G., ``The binary black-hole
  dynamics at the third-and-a-half post-Newtonian order in the ADM-formalism'',
  {\em Phys. Rev. D}, {\bf 68}, 044004, (2003).
  {\small[\href{http://arxiv.org/abs/gr-qc/0305048}{{gr-qc/0305048}}]}.

\bibitem{ALeTiec2012}
Le~Tiec, A., Barausse, E.  and Buonanno, A., ``Gravitational Self-Force
  Correction to the Binding Energy of Compact Binary Systems'', {\em Phys. Rev.
  Lett.}, {\bf 108}, 131103, (2012).
  {\small[\href{http://arxiv.org/abs/1111.5609 [gr-qc]}{{1111.5609 [gr-qc]}}]}.

\bibitem{LeTiec2012}
Le~Tiec, A., Blanchet, L.  and Whiting, B.F., ``The First Law of Binary Black
  Hole Mechanics in General Relativity and Post-Newtonian Theory'', {\em Phys.
  Rev. D}, {\bf 85}, 064039, (2012).
  {\small[\href{http://arxiv.org/abs/1111.5378 [gr-qc]}{{1111.5378 [gr-qc]}}]}.

\bibitem{LeTiec2011}
Le~Tiec, A., Mroue, A.H., Barack, L., Buonanno, A., Pfeiffer, H.P., Sago, N.
  and Taracchini, A., ``Periastron Advance in Black Hole Binaries'', {\em Phys.
  Rev. Lett.}, {\bf 107}, 141101, (2011).
  {\small[\href{http://arxiv.org/abs/1106.3278 [gr-qc]}{{1106.3278 [gr-qc]}}]}.

\bibitem{MacDonald:2012mp}
MacDonald, Ilana, Mroue, Abdul~H., Pfeiffer, Harald~P., Boyle, Michael, Kidder,
  Lawrence~E.  {et~al.}, ``{Suitability of hybrid gravitational waveforms for
  unequal-mass binaries}'', (2012).
  {\small[\href{http://arxiv.org/abs/1210.3007}{{arXiv:1210.3007
  {\small[gr-qc]}}}]}.

\bibitem{Nagar2011}
Nagar, A., ``Effective one body Hamiltonian of two spinning black-holes with
  next-to-next-to-leading order spin-orbit coupling'', {\em Phys. Rev. D}, {\bf
  84}, 084028, (2011). {\small[\href{http://arxiv.org/abs/1106.4349
  [gr-qc]}{{1106.4349 [gr-qc]}}]}.

\bibitem{Nissanke2005}
Nissanke, S.  and Blanchet, L., ``Gravitational radiation reaction in the
  equations of motion of compact binaries to 3.5 post-Newtonian order'', {\em
  Class. Quant. Grav.}, {\bf 22}, 1007, (2005).
  {\small[\href{http://arxiv.org/abs/gr-qc/0412018}{{gr-qc/0412018}}]}.

\bibitem{Pan2011}
Pan, Y., Buonanno, A., Boyle, M., Buchman, L.T., Kidder, L.E., Pfeiffer, H.P.
  and Scheel, M.A., ``Inspiral-merger-ringdown multipolar waveforms of
  nonspinning black-hole binaries using the effective-one-body formalism'',
  {\em Phys. Rev. D}, {\bf 84}, 124052, (2011).
  {\small[\href{http://arxiv.org/abs/1106.1021 [gr-qc]}{{1106.1021 [gr-qc]}}]}.

\bibitem{Pan2010}
Pan, Y., Buonanno, A., Buchman, L.T., Chu, T., Kidder, L.E., Pfeiffer, H.P.
  and Scheel, M.A., ``Effective-one-body waveforms calibrated to numerical
  relativity simulations: coalescence of non-precessing, spinning, equal-mass
  black holes'', {\em Phys. Rev. D}, {\bf 81}, 084041, (2010).
  {\small[\href{http://arxiv.org/abs/0912.3466 [gr-qc]}{{0912.3466 [gr-qc]}}]}.

\bibitem{YPan2011}
Pan, Y., Buonanno, A., Fujita, R., Racine, E.  and Tagoshi, H.,
  ``Post-Newtonian factorized multipolar waveforms for spinning, non-precessing
  black-hole binaries'', {\em Phys. Rev. D}, {\bf 83}, 064003, (2011).
  {\small[\href{http://arxiv.org/abs/1006.0431 [gr-qc]}{{1006.0431 [gr-qc]}}]}.

\bibitem{Pan2008}
Pan, Y.~\textit{et al.}, ``A data-analysis driven comparison of analytic and
  numerical coalescing binary waveforms: Nonspinning case'', {\em Phys. Rev.
  D}, {\bf 77}, 024014, (2008). {\small[\href{http://arxiv.org/abs/0704.1964
  [gr-qc]}{{0704.1964 [gr-qc]}}]}.

\bibitem{Will2000}
Pati, M.E.  and Will, C.M., ``Post-Newtonian gravitational radiation and
  equations of motion via direct integration of the relaxed Einstein equations.
  I: Foundations'', {\em Phys. Rev. D}, {\bf 62}, 124015, (2000).
  {\small[\href{http://arxiv.org/abs/gr-qc/0007087}{{gr-qc/0007087}}]}.

\bibitem{Pati2002}
Pati, M.E.  and Will, C.M., ``Post-Newtonian gravitational radiation and
  equations of motion via direct integration of the relaxed Einstein equations.
  II: Two-body equations of motion to second post-Newtonian order, and
  radiation-reaction to 3.5 post-Newton'', {\em Phys. Rev. D}, {\bf 65},
  104008, (2002).
  {\small[\href{http://arxiv.org/abs/gr-qc/0201001}{{gr-qc/0201001}}]}.

\bibitem{Poisson1993}
Poisson, E., ``Gravitational radiation from a particle in circular orbit around
  a black hole. I. Analytic results for the nonrotating case'', {\em Phys. Rev.
  D}, {\bf 47}, 1497--1510, (1993).

\bibitem{Pretorius2005}
Pretorius, F., ``Evolution of Binary Black Hole Spacetimes'', {\em Phys. Rev.
  Lett.}, {\bf 95}, 121101, (2005).
  {\small[\href{http://arxiv.org/abs/gr-qc/0507014}{{gr-qc/0507014}}]}.

\bibitem{Pretorius2007}
Pretorius, F., ``Binary Black Hole Coalescence'', in Colpi, M.~{\it et al.},
  ed., {\em Relativistic Objects in Compact Binaries: From Birth to
  Coalescense}. Springer Verlags, Canopus Publishing Limited, (2007).
  {\small[\href{http://arxiv.org/abs/arXiv:0710.1338 [gr-qc]}{{arXiv:0710.1338
  [gr-qc]}}]}.

\bibitem{Price1994}
Price, R.H.  and Pullin, J., ``Colliding black holes: The Close limit'', {\em
  Phys. Rev. Lett.}, {\bf 72}, 3297, (1994).
  {\small[\href{http://arxiv.org/abs/gr-qc/9402039}{{gr-qc/9402039}}]}.

\bibitem{Sasaki1994}
Sasaki, M., ``Post-Newtonian Expansion of the Ingoing-Wave Regge-Wheeler
  Function'', {\em Prog. Theor. Phys}, {\bf 92}, 17--36, (1994).

\bibitem{Sasaki2003}
Sasaki, M.  and Tagoshi, H., ``Analytic black hole perturbation approach to
  gravitational radiation'', {\em Living Rev. Rel.}, {\bf 6}, 6, (2003).
  {\small[\href{http://arxiv.org/abs/gr-qc/0306120}{{gr-qc/0306120}}]}.

\bibitem{Scheel2009}
Scheel, M.A., Boyle, M., Chu, T., Kidder, L.E., Matthews, K.D.  and Pfeiffer,
  H.P., ``High-accuracy waveforms for binary black hole inspiral, merger, and
  ringdown'', {\em Phys. Rev. D}, {\bf 79}, 024003, (2009).
  {\small[\href{http://arxiv.org/abs/0810.1767 [gr-qc]}{{0810.1767 [gr-qc]}}]}.

\bibitem{Tagoshi1994}
Tagoshi, H.  and Sasaki, M., ``Post-Newtonian Expansion of Gravitational Waves
  from a Particle in Circular Orbit around a Schwarzschild Black Hole'', {\em
  Prog. Theor. Phys}, {\bf 92}, 745--771, (1994).
  {\small[\href{http://arxiv.org/abs/gr-qc/9405062}{{gr-qc/9405062}}]}.

\bibitem{Tanaka1996}
Tanaka, T., Tagoshi, H.  and Sasaki, M., ``Gravitational Waves by a Particle in
  Circular Orbit around a Schwarzschild Black Hole'', {\em Prog. Theor. Phys},
  {\bf 96}, 1087--1101, (1996).
  {\small[\href{http://arxiv.org/abs/gr-qc/9701050}{{gr-qc/9701050}}]}.

\bibitem{Taracchini2012}
Taracchini, A., Pan, Y., Buonanno, A., Barausse, E., Boyle, M., Chu, T.,
  Lovelace, G.  and Pfeiffer, H.P.~{\it et al.}, ``Prototype effective-one-body
  model for nonprecessing spinning inspiral-merger-ringdown waveforms'', {\em
  Phys. Rev. D}, {\bf 86}, 024011, (2012).
  {\small[\href{http://arxiv.org/abs/1202.0790 [gr-qc]}{{1202.0790 [gr-qc]}}]}.

\bibitem{Thorne1980}
Thorne, K.S., ``Multipole expansions of gravitational radiation'', {\em Rev.
  Mod. Phys.}, {\bf 52}, 299--340, (1980).

\bibitem{Wagoner1976}
Wagoner, R.V.  and Will, C.M., ``Post-Newtonian gravitational radiation from
  orbiting point masses'', {\em Astrophys. J.}, {\bf 210}, 764--775, (1976).

\bibitem{Will1999}
Will, C.M., ``Generation of post-Newtonian gravitational radiation via direct
  integration of the relaxed Einstein equations'', {\em Prog. Theor. Phys.
  Suppl.}, {\bf 136}, 158, (1999).
  {\small[\href{http://arxiv.org/abs/gr-qc/9910057}{{gr-qc/9910057}}]}.

\bibitem{Will1996}
Will, C.M.  and Wiseman, A.G., ``Gravitational radiation from compact binary
  systems: gravitational waveforms and energy loss to second post-Newtonian
  order'', {\em Phys. Rev. D}, {\bf 54}, 4813, (1996).
  {\small[\href{http://arxiv.org/abs/gr-qc/9608012}{{gr-qc/9608012}}]}.

\bibitem{Wiseman1993}
Wiseman, A.G., ``Coalescing Binary Systems Of Compact Objects To
  (Post)Newtonian5/2 Order.4v: The Gravitational Wave Tail'', {\em Phys. Rev.
  D}, {\bf 48}, 4757, (1993).

\bibitem{Yunes2010}
Yunes, N., Buonanno, A., Hughes, S.A., Coleman~Miller, M.  and Pan, Y.,
  ``Modeling Extreme Mass Ratio Inspirals within the Effective-One-Body
  Approach'', {\em Phys. Rev. Lett.}, {\bf 104}, 091102, (2010).
  {\small[\href{http://arxiv.org/abs/0909.4263 [gr-qc]}{{0909.4263 [gr-qc]}}]}.

\bibitem{Yunes2011}
Yunes, N., Buonanno, A., Hughes, Pan, Y., Barausse, E., Miller, M.C.  and
  Throwe, W., ``Extreme Mass-Ratio Inspirals in the Effective-One-Body
  Approach: Quasi-Circular, Equatorial Orbits around a Spinning Black Hole'',
  {\em Phys. Rev. D}, {\bf 83}, 044044, (2011).
  {\small[\href{http://arxiv.org/abs/1009.6013 [gr-qc]}{{1009.6013 [gr-qc]}}]}.

\end{thebibliography}

\end{document}